\documentclass[12pt]{article}

\usepackage{amsmath,mathtools,amsfonts}
\usepackage{amssymb,xypic}
\usepackage{bm}
\usepackage[dvips]{graphicx}
\usepackage[dvips]{graphics}
\usepackage[all]{xy}
\usepackage{mathrsfs}
\usepackage{cite}

\newcommand{\sectiono}[1]{\section{#1}\setcounter{equation}{0}}

\newcommand{\bd}[1]{\boldsymbol{#1}}
\newcommand{\id}{\mathbb{I}}

\newcommand{\NSNS}{NS\textrm{-}NS}
\newcommand{\RNS}{R\textrm{-}NS}
\newcommand{\NSR}{NS\textrm{-}R}
\newcommand{\RR}{R\textrm{-}R}
\newcommand{\sst}{(s,\bar{s},t)}
\newcommand{\ssb}{(s,\bar{s})}
\newcommand{\llb}{(\bd{\lambda}+\bar{\bd{\lambda}})}

\begin{document}

\baselineskip=17pt

\begin{titlepage}
\rightline{\tt YITP-21-58}
%\rightline\today
\begin{center}
\vskip 2.5cm
{\Large \bf {Type II superstring field theory revisited}}
\vskip 1.0cm
{\large {Hiroshi Kunitomo}}
\vskip 1.0cm
{\it {Center for Gravitational Physics}}, 
{\it {Yukawa Institute for Theoretical Physics}}\\
{\it {Kyoto University}},
{\it {Kitashirakawa Oiwakecho, Sakyo-ku, Kyoto 606-8502, Japan}}\\
kunitomo@yukawa.kyoto-u.ac.jp

\vskip 2.0cm

{\bf Abstract}
\end{center}
\noindent

We reconstruct a complete type II superstring field theory with $L_\infty$ structure 
in a symmetric way concerning the left- and right-moving sectors. Based on the new construction, 
we show again that the tree-level S-matrix agrees with that obtained using the first-quantization 
method. Not only is this a simple and elegant reconstruction, but it also enables the action 
to be mapped to that in the WZW-like superstring field theory, which has
not yet been constructed and fills the only gap in the WZW-like formulation.

\end{titlepage}

\tableofcontents

\newpage

\sectiono{Introduction}

Recent significant developments in the superstring field theory 
\cite{Berkovits:1995ab,Berkovits:2004xh,Matsunaga:2014wpa,Erler:2015rra,
Kunitomo:2015usa,Matsunaga:2015kra,Erler:2017onq,
Jurco:2013qra,Erler:2013xta,Erler:2014eba,Erler:2015lya,
Erler:2016ybs,Matsunaga:2016zsu,Kunitomo:2019glq,
Kunitomo:2019kwk,Sen:2015hha,Sen:2015uaa,Konopka:2016grr}
have ended long-term stagnation and made it possible to construct classical gauge-invariant 
actions for open \cite{Erler:2016ybs,Kunitomo:2015usa,Konopka:2016grr}, 
heterotic \cite{Kunitomo:2019glq,Sen:2015hha,Sen:2015uaa}, 
and type II \cite{Kunitomo:2019kwk,Sen:2015hha,Sen:2015uaa} superstring theories. It is based on 
three complementary construction methods:
the formulation with the homotopy algebra structure\cite{Erler:2013xta,Erler:2014eba,Erler:2015lya,
Erler:2016ybs,Kunitomo:2019glq,Kunitomo:2019kwk}, 
the Wess-Zumino-Witten (WZW) -like formulation \cite{Berkovits:1995ab,Berkovits:2004xh,
Matsunaga:2014wpa,Erler:2015rra,Kunitomo:2015usa,Matsunaga:2015kra,Erler:2017onq,
Matsunaga:2016zsu,Kunitomo:2019glq,Kunitomo:2019kwk}, 
and the formulation accompanied with an extra free field 
\cite{Konopka:2016grr,Sen:2015hha,Sen:2015uaa}.
The only exception is the complete WZW-like action for the type II superstring field theory, 
which has not yet been constructed.
One of the purposes of this paper is to give it 
% a complete WZW-like action for the type II superstring 
and fills a gap in the WZW-like formulation.

In the previous papers \cite{Kunitomo:2019glq,Kunitomo:2019kwk}, 
we have attempted to construct complete WZW-like actions, including 
all sectors representing the space-time bosons and fermions, for the heterotic 
and the type II superstring field theories by mapping from those of the formulation 
with the $L_\infty$ algebra method. This construction worked well for the heterotic
string theory but not for the type II superstring theory. 
The reason is that we have taken the left-right asymmetric method
for constructing the type II string field theory with an $L_\infty$ structure. 
In this case, we could only construct an action that is a hybrid 
of the WZW-like action and the action with an $L_\infty$ structure, which we
called the half-WZW-like action. 
Therefore, we first revisit the formulation based on $L_\infty$ structure
and provide an alternative to the asymmetric method, the symmetric way 
to construct the action for the type II superstring field theory. 
It is another purpose of this paper.

The paper is organized as follows. In section \ref{SST with L}, we summarize 
how to construct a gauge-invariant action for the type II superstring field theory based on 
an $L_\infty$ structure. The string field is suitably constrained and the gauge-invariant 
action is written by using the string products satisfying the $L_\infty$ algebra relations.
We propose in section \ref{sym const} a new construction method of the string products 
with cyclic $L_\infty$ structure symmetric with respect to the left- and right-moving sectors.
After introducing the coalgebra representation, we suppose $L_\infty$ algebra using 
the string products with fixed cyclic Ramond numbers. The significant point, which is different
from the other cases, is that their RR output part is including the picture-changing operator (PCO)
explicitly. This breaks the cyclicity of the algebra but instead suitably decomposes it into 
three commutative $L_\infty$ algebras (rather than four), two constraint $L_\infty$ algebras 
and a dynamical $L_\infty$ algebra.
A similarity transformation allows us to transform dynamical $L_\infty$ algebra 
into the one used for writing down the action and the other two into the constraints 
that impose that the first algebra closes in the small Hilbert space.
In order to explicitly construct such cyclic string products, 
we give differential equations for their generating functional.
These differential equations recursively determine the string products with respect to the number 
of input string fields, with bosonic products as the initial condition.
We show that the new action correctly derives the same tree-level physical S-matrix as that 
calculated using the first-quantization method in section \ref{sec S-matrix}.
The proof is based on the homological perturbation theory (HPT), which provides the explicit 
form of the tree-level S-matrix in closed form and makes it possible to demonstrate 
the agreement.
In section  \ref{sec WZW}, we write down a complete WZW-like action for the type II 
superstring field theory, which we could not construct previously.
After summarizing how the NS-NS action with $L_\infty$ structure was rewritten 
as a WZW-like action through the map between string fields in two formulations, 
we extend it to all four sectors. It fills the gap in the WZW-like formulation
and should be significant to deepen the understanding of the superstring field theory.
Section \ref{concl} is devoted to the conclusion and discussion.
Finally, it contains two appendices. In Appendix A, we define a projected commutator 
that plays a significant role in constructing the cyclic products with the $L_\infty$ 
structure. Appendix B contains the $s$ and $\bar{s}$ expansions of the generating functionals of 
cyclic three-string products and corresponding gauge products to help for understanding 
the flow how they are recursively determined.

\sectiono{Type II superstring field theory with $L_\infty$ structure}\label{SST with L}

Let us recall how the type II superstring field theory %, including all the four sectors, 
with an $L_\infty$ structure has been constructed \cite{Erler:2014eba,
Kunitomo:2019kwk}. The first-quantized Hilbert space, $\mathcal{H}$\,, of type II superstring
is composed of four sectors corresponding to the combination of the NS and Ramond boundary 
conditions for the left- and right-moving fermionic coordinates: 
\begin{equation}
 \mathcal{H}\ =\ \mathcal{H}_{\NSNS} + \mathcal{H}_{\RNS} + \mathcal{H}_{\NSR} + \mathcal{H}_{\RR}\,.
\end{equation}
Correspondingly, the type II superstring field $\Phi$ has four components:
\begin{equation}
 \Phi\ =\ \Phi_{\NSNS} + \Phi_{\RNS} + \Phi_{\NSR} + \Phi_{\RR} \in \mathcal{H}\,,
\end{equation}
which is Grassmann even\footnote{Here, for consistent Grassmann property, 
we should generally assume that $\Phi$ is GSO projected.} 
and has ghost number 2.
We take the picture number of each component as $(-1,-1)$\,, $(-1/2,-1)$\,, $(-1,-1/2)$ 
and $(-1/2,-1/2)$\,, respectively. 
The NS-NS and R-R components, $\Phi_{\NSNS}$ and 
$\Phi_{\RR}$\,, represent space-time bosons, and the R-NS and NS-R components, $\Phi_{\RNS}$ and 
$\Phi_{\NSR}$\,, represent space-time fermions.
The string field $\Phi$ is restricted by the closed string constraints
\begin{equation}
 b_0^-\Phi\ =\ L_0^-\Phi\ =\ 0\,,
\label{closed string constraints}
\end{equation}
with $b_0^\pm=b_0\pm\bar{b}_0,\ L_0^\pm=L_0\pm\bar{L}_0$\,, and $c_0^\pm=(c_0\pm\bar{c}_0)/2$\,.
We call the Hilbert space restricted by constraints (\ref{closed string constraints}) the closed string Hilbert space.
In addition, it is necessary to impose an extra constraint by introducing a projection operator 
$\mathcal{P}_{\mathcal{G}}=\mathcal{G}\mathcal{G}^{-1}$ with
\begin{equation}
\begin{split}
  \mathcal{G}\ =&\ \pi^{(0,0)} + X\pi^{(1,0)} + \bar{X}\pi^{(0,1)} + X\bar{X}\pi^{(1,1)}\,,\\
 \mathcal{G}^{-1}\ =&\ \pi^{(0,0)} + Y\pi^{(1,0)} + \bar{Y}\pi^{(0,1)} + Y\bar{Y}\pi^{(1,1)}\,.
\end{split}
\end{equation}
Here, $\pi^{(a,b)}\ (a,b=0,1)$ is a projection operator onto a component
\begin{equation}
 \pi^{(0,0)}\Phi\ =\ \Phi_{\NSNS}\,,\
 \pi^{(1,0)}\Phi\ =\ \Phi_{\RNS}\,,\
 \pi^{(0,1)}\Phi\ =\ \Phi_{\NSR}\,,\
 \pi^{(1,1)}\Phi\ =\ \Phi_{\RR}\,,
\end{equation}
and, we take the PCO's as
\begin{alignat}{2}
 X\ =&\ -\delta(\beta_0)G + \frac{1}{2}(\gamma_0\delta(\beta_0)+\delta(\beta_0)\gamma_0)b_0^+\,,&\qquad
 Y\ =&\ -2\frac{G}{L_0^+}\delta(\gamma_0)\,,\\
 \bar{X}\ =&\ -\delta(\bar{\beta}_0)\bar{G} 
+ \frac{1}{2}(\bar{\gamma}_0\delta(\bar{\beta}_0)+\delta(\bar{\beta}_0)\bar{\gamma}_0)b_0^+\,,&\qquad
 \bar{Y}\ =&\ -2\frac{\bar{G}}{L_0^+}\delta(\bar{\gamma}_0)\,.
\end{alignat}
Note that $\mathcal{G}$ satisfies
\begin{equation}
 \mathcal{G}\mathcal{G}^{-1}\mathcal{G}\ =\ \mathcal{G}\,,\qquad
 \mathcal{G}^{-1}\mathcal{G}\mathcal{G}^{-1}\ =\ \mathcal{G}^{-1},\qquad
[Q,\mathcal{G}]\ =\ 0\,,
\end{equation}
and is almost exact in the large Hilbert space $\mathcal{H}_l$:
\begin{equation}
\mathcal{G}\ =\ \pi^{(0,0)} + [Q\,, \mathfrak{S}]\,,\qquad
 \mathfrak{S}\ =\
\Xi\pi^{(1,0)}+\bar{\Xi}\pi^{(0,1)} 
+ \frac{1}{2}(\Xi\bar{X}+X\bar{\Xi})\pi^{(1,1)}\,,
\end{equation}
with\footnote{For notational simplicity, we denote the zero modes of fermionic ghosts 
$(\xi(z),\eta(z))$ and $(\bar{\xi}(\bar{z}),\bar{\eta}(\bar{z}))$ as $(\xi,\eta)$ 
and $(\bar{\xi},\bar{\eta})$\,, respectively.}
\begin{align}
 \Xi\ =&\ \xi + (\Theta(\beta_0)\eta\xi - \xi)P_{-3/2}  + (\xi\eta\Theta(\beta_0)-\xi)P_{-1/2}\,,\\
 \bar{\Xi}\ =&\ \bar{\xi} + (\Theta(\bar{\beta}_0)\bar{\eta}\bar{\xi} - \bar{\xi})P_{-3/2} 
+ (\bar{\xi}\bar{\eta}\Theta(\beta_0)-\bar{\xi})P_{-1/2}\,.
\end{align}
The operator $P_{-3/2}$ and $P_{-1/2}$ are the projector onto the states with ghost number
$-3/2$ and $-1/2$\,, respectively. The BRST operator consistently acts on the string field
restricted by (\ref{additional constraint}) thanks to the property: 
$\mathcal{P}_{\mathcal{G}}Q\mathcal{P}_{\mathcal{G}}=Q\mathcal{P}_{\mathcal{G}}$\,.
Then, the last constraint 
\begin{equation}
\mathcal{P}_{\mathcal{G}}\Phi\ =\ \Phi
\label{additional constraint}
\end{equation}
restricts the dependence of the ghost zero modes of each component of the string field as
\begin{subequations}
 \begin{align}
 \Phi_{\NSNS}\ =&\ \phi_{\NSNS} - c_0^+\psi_{\NSNS}\,,\\
 \Phi_{\RNS}\ =&\ \phi_{\RNS} - \frac{1}{2}(\gamma_0+2c_0^+G)\psi_{\RNS}\,,\\
 \Phi_{\NSR}\ =&\ \phi_{\NSR} - \frac{1}{2}(\bar{\gamma}_0+2c_0^+\bar{G})\psi_{\NSR}\,,\\
 \Phi_{\RR}\ =&\ \phi_{\RR} - \frac{1}{2}(\gamma_0\bar{G}-\bar{\gamma}_0G + 2c_0^+G\bar{G})\psi_{\RR}\,.
\end{align}
\end{subequations}
We call the Hilbert space further restricted by (\ref{additional constraint}) as the restricted Hilbert space, 
$\mathcal{H}^{res}$\,.
The ghost-zero-mode independent components $\phi_i$ ($\psi_i$) $(i=\NSNS,\RNS,\NSR,\RR)$
are Grassmann even (odd) and correspond to the fields (anti-fields) in the gauge-fixed basis
when they are quantized by the Batalin-Vilkovisky (BV) formalism. 
The simplest and practical gauge is obtained by setting $\psi=0$\,, which we call the Siegel-Ramond (SR) gauge
and denote its Hilbert space as $\mathcal{H}_{SR}$\,.

A natural symplectic form in the closed string Hilbert space is defined by using the BPZ-inner 
product as
\begin{equation}
 \omega(\Phi_1\,,\Phi_2)\ =\ (-1)^{|\Phi_1|}\langle\Phi_1|c_0^-|\Phi_2\rangle\,,
\end{equation}
where $\langle\Phi|$ is the BPZ conjugate of $|\Phi\rangle$\,. The symbol $|\Phi|$ denotes
the Grassmann property of string field $\Phi$: $|\Phi|=0$ or 1 if $\Phi$ is Grassmann even or odd, respectively.
A natural symplectic form in the restricted Hilbert space $\Omega$ is defined by using $\omega$ as
\begin{equation}
 \Omega(\Phi_1\,,\Phi_2)\ =\ \omega(\Phi_1\,,\mathcal{G}^{-1}\Phi_2)\,.
\end{equation} 
Due to the asymmetry of the inner product among sectors, it is nontrivial to make the $L_\infty$
algebra being cyclic across all the sectors.
For later use, it is also convenient to introduce the symplectic form $\omega_l$
in the large Hilbert space. It is related to $\omega$ as
\begin{equation}
 \omega_l(\xi\bar{\xi} \Phi_1\,,\Phi_2)\ =\ \omega(\Phi_1\,, \Phi_2)
\end{equation}
if we embed $\Phi_1,\Phi_2\in\mathcal{H}$ into $\mathcal{H}_l$ as the fields
satisfying the constraint $\eta\Phi_i=\bar{\eta}\Phi_i=0\ (i=1,2)$\,. 
We also use the bilinear map representation of symplectic forms defined by
\begin{equation}
 \begin{alignedat}{6}
\langle\omega_l|\,:\,
\mathcal{H}_l&\otimes\mathcal{H}_l\qquad & 
&\ {\longrightarrow}&\qquad &\mathbb{C}
\\
&\ \rotatebox{90}{$\in$} & & &  &\rotatebox{90}{$\in$}
\\
 \Phi_1&\otimes\Phi_2\qquad 
& &\longmapsto&\qquad \omega_l(\Phi&_1\,,\Phi_2)\,,
\end{alignedat}
\end{equation} 
and
\begin{equation}
 \langle\omega_s|\ =\ \langle\omega_l|\xi\bar{\xi}\otimes\id\,,\qquad
 \langle\Omega|\ =\ \langle\omega_l|\xi\bar{\xi}\otimes\mathcal{G}^{-1}\,.
\end{equation}

The action of the type II superstring field theory is defined using the string products $L_n$
mapping $n$ string fields to a string field as
\begin{alignat}{3}
 L_n\ :\qquad\qquad 
&\ (\mathcal{H}^{res})^{\wedge n} &\qquad &\ \longrightarrow&\qquad  &\ \mathcal{H}^{res}\,,\qquad
(n\ge1)\,,
\nonumber\\
&\ \ \ \  \text{\rotatebox[origin=c]{90}{$\in$}}&\qquad &\ &\qquad &\ \ \text{\rotatebox[origin=c]{90}{$\in$}}\\
\ \ \  \Phi_1&\wedge \ \cdots\wedge\Phi_n &\qquad &\ \longmapsto&\qquad  L_n(\Phi_1&\,,\ \cdots\,, \Phi_n)\,.
\nonumber
\end{alignat}
We identify the one-string product as the BRST operator $L_1=Q$\,, and
each $(\mathcal{H}^{res})^{\wedge n}$ $(n\ge2)$ is the symmetrized tensor product of 
$\mathcal{H}^{res}$ whose element, 
$\Phi_1\wedge\cdots\wedge\Phi_n\in(\mathcal{H}^{res})^{\wedge n}$\,, is defined by
\begin{equation}
\Phi_1\wedge\cdots\wedge\Phi_n\ =\ \sum_{\sigma}(-1)^{\epsilon(\sigma)}
\Phi_{\sigma(1)}\otimes\cdots\otimes\Phi_{\sigma(n)}\,,\qquad
\Phi_i\in\mathcal{H}^{res}\,,
\end{equation}
where $\sigma$ and $\epsilon(\sigma)$ denote all the permutations of $\{1,\cdots,n\}$
and the sign factor coming from the exchange $\{\Phi_1,\cdots,\Phi_n\}$ to 
$\{\Phi_{\sigma(1)},\cdots,\Phi_{\sigma(n)}\}$\,. 
Note that the string field $L_n(\Phi_1,\cdots,\Phi_n)$
must satisfy the constraints (\ref{closed string constraints}) and (\ref{additional constraint}):
\begin{align}
\begin{split} 
&\ b_0^-L_n(\Phi_1\,,\cdots\,,\Phi_n)\ =\ L_0^-L_n(\Phi_1\,,\cdots\,,\Phi_n)\ =\ 0\,,\\
&\ \mathcal{P}_{\mathcal{G}}L_n(\Phi_1\,,\cdots\,,\Phi_n)\ =\ L_n(\Phi_1\,,\cdots\,,\Phi_n)\,.
\end{split}
\end{align}
If the string products are equipped with the cyclic $L_\infty$ structure, that is, satisfy 
the $L_\infty$ relations
\begin{equation}
  \sum_{\sigma}\sum_{m=1}^n(-1)^{\epsilon(\sigma)}
\frac{1}{m!(n-m)!}L_{n-m+1}(L_m(\Phi_{\sigma(1)}\,,\cdots\,,\Phi_{\sigma(m)})\,, \Phi_{\sigma(m+1)}\,,
\cdots\,,\Phi_{\sigma(n)})\ =\ 0\,,
\label{L infinity relations}
 \end{equation}
and the cyclicity condition, %with respect to the symplectic form $\Omega$\,:
\begin{equation}
 \Omega(\Phi_1\,, L_n(\Phi_2\,,\cdots\,,\Phi_{n+1}))\ 
=\ -(-1)^{|\Phi_1|}\Omega(L_n(\Phi_1\,,\cdots\,,\Phi_n)\,,\Phi_{n+1})\,,
\label{cyclicity condition}
\end{equation}
the action of the type II superstring field theory is given by
\begin{equation}
 I\ =\ \sum_{n=0}^\infty\frac{1}{(n+2)!}\Omega(\Phi\,,L_{n+1}(\underbrace{\Phi\,,\cdots\,,\Phi}_{n+1}))\,,
\label{action}
\end{equation}
which is invariant under the gauge transformation
\begin{equation}
 \delta\Phi\ =\ \sum_{n=0}^\infty\frac{1}{n!}L_{n+1}(\underbrace{\Phi\,,\cdots\,,\Phi}_n\,, \Lambda)\,.
\end{equation}
The gauge parameter $\Lambda$ has also four components:
$\Lambda=\Lambda_{\NSNS} + \Lambda_{\RNS} + \Lambda_{\NSR} + \Lambda_{\RR}\in\mathcal{H}$\,.
A set of string products satisfying (\ref{L infinity relations}) and (\ref{cyclicity condition}) 
defines a cyclic $L_\infty$ algebra $(\mathcal{H}^{res},\Omega,\{L_n\})$\,.
The nontrivial task is to provide a prescription that gives these string products.

\sectiono{Symmetric construction of string products}\label{sym const}

In the previous paper \cite{Kunitomo:2019kwk}, we gave a prescription 
to construct a set of string products with an $L_\infty$ structure required
by complete action of the type II superstring field theory by extending 
the asymmetric construction proposed in \cite{Erler:2014eba}.
It is sufficient for defining an action and gauge transformation
% of type II superstring field theory 
but is slightly complicated and could not be mapped to those in the WZW-like 
formulation. In this paper, we propose another prescription by extending
the symmetric construction in \cite{Erler:2014eba}.

\subsection{$L_\infty$ triplet}

We use the coalgebra representation. The coderivation $\bd{L}=\sum_{n=0}^\infty\bd{L}_n$ is
defined as an operator acting on symmetrized tensor algebra
\begin{equation}
 \mathcal{SH}\ =\ \mathcal{H}^{\wedge 0}\oplus\mathcal{H}\oplus\mathcal{H}^{\wedge 2}\oplus\cdots\,,
\end{equation}
by\footnote{The identity operator $\id_n$ acting on $\mathcal{H}^{\wedge n}$ is given by
$\id_n=\frac{1}{n!}\underbrace{\id\wedge\cdots\wedge\id}_n 
=\underbrace{\id\otimes\cdots\otimes\id}_n.$}
\begin{alignat}{2}
 L_n\Phi_1\wedge\cdots\wedge\Phi_m\ =&\ 0\,,\qquad & &\textrm{for}\ m<n\,,\\
 L_n\Phi_1\wedge\cdots\wedge\Phi_m\ =&\ L_n(\Phi_1\,,\cdots\,,\Phi_n)\,,\qquad & &\textrm{for}\ m=n\,,\\
 L_n\Phi_1\wedge\cdots\wedge\Phi_m\ =&\ (L_n\wedge\id_{m-n})\Phi_1\wedge\cdots\wedge\Phi_m\,,
\qquad & &\textrm{for}\ m<n\,.
\end{alignat}
We denote the projection onto $\mathcal{H}^{\wedge n}$ as $\pi_n$\,:
$\pi_n\mathcal{SH}=\mathcal{H}^{\wedge n}$\,.
The string products satisfying $L_\infty$ relation (\ref{L infinity relations}) is given by
a nilpotent odd coderivation, $\bd{L}=\sum_{n=1}^\infty\bd{L}_n$ satisfying\footnote{
In this paper the square bracket $[\cdot,\cdot]$ denotes the graded commutator.}
 $[\bd{L},\bd{L}]=0$\,. 
The basic building block of our construction is the odd coderivation 
\begin{equation}
 \bd{B}\ =\ \sum_{p,r=0}^\infty\sum_{\bar{p},\bar{r}=0}^\infty 
\bd{B}^{(p,\bar{p})}_{p+r+1,\bar{p}+\bar{r}+1}\mid^{(2r,2\bar{r})}\,,
\label{coderivation B}
\end{equation}
acting on the symmetrized tensor algebra, $\mathcal{SH}_l$\,, generated by the large Hilbert space.
The superscripts $p$ $(\bar{p})$  denotes the left- (right-) moving picture number of the string product 
itself and $2r$ $(2\bar{r})$ denotes the left- (right-) moving cyclic Ramond number defined by
\begin{center}
 Cyclic Ramond No. $=$ No. of Ramond inputs $+$ No. of Ramond outputs.
\end{center}
We used a matrix notation
describing the string products as a diagonal matrix \cite{Erler:2014eba}:
%the subscript representing the number of inputs was duplicated, 
%and described the string product as a diagonal matrix:
\begin{equation}
 \bd{B}^{(p,\bar{p})}_{p+r+1,\bar{p}+\bar{r}+1}\mid^{(2r,2\bar{r})}\ =\ 
\delta_{p+r,\bar{p}+\bar{r}}\bd{B}^{(p,\bar{p})}_{p+r+1}\mid^{(2r,2\bar{r})}\,.
\end{equation}
%The prameter $t$ is counting the total picture number and will be set equal to 1 in the end.
The coderivation $\bd{B}$ can be decomposed into four components according to 
the sector of outputs as: %$\pi_1\bd{B}(t)=(\pi^{(0,0)}_1+\pi^{(1,0)}_1+\pi^{(1,0)}_1+\pi^{(1,0)}_1)\bd{B}$\,:
\begin{subequations} 
 \begin{align}
\pi^{(0,0)}_1\bd{B}\ =&\  \sum_{p,r=0}^\infty\sum_{\bar{p},\bar{r}=0}^\infty 
\pi_1\bd{B}^{(p,\bar{p})}_{p+r+1,\bar{p}+\bar{r}+1}\mid^{(2r,2\bar{r})}_{(2r,2\bar{r})}\,,
\label{B11}\\
%%%%%%%%%%%%%%%%%%%
 \pi^{(1,0)}_1\bd{B}\ =&\ \sum_{p,r=0}^\infty\sum_{\bar{p},\bar{r}=0}^\infty
\pi_1\bd{B}^{(p,\bar{p})}_{p+r+1,\bar{p}+\bar{r}+1}\mid^{(2r,2\bar{r})}_{(2r-2,2\bar{r})}\,,
\label{B21}\\
%%%%%%%%%%%%%%%%%%%
 \pi^{(0,1)}_1\bd{B}\ =&\ \sum_{p,r=0}^\infty\sum_{\bar{p},\bar{r}=0}^\infty 
\pi_1\bd{B}^{(p,\bar{p})}_{p+r+1,\bar{p}+\bar{r}+1}\mid^{(2r,2\bar{r})}_{(2r,2\bar{r}-2)}\,,
\label{B12}\\
%%%%%%%%%%%%%%%%%%%
 \pi^{(1,1)}_1\bd{B}\ =&\ \sum_{p,r=0}^\infty\sum_{\bar{p},\bar{r}=0}^\infty 
\pi_1\bd{B}^{(p,\bar{p})}_{p+r+1,\bar{p}+\bar{r}+1}\mid^{(2r,2\bar{r})}_{(2r-2,2\bar{r}-2)},
\label{B22}
\end{align}
\end{subequations}
where $\pi_1^{(a,b)}=\pi^{(a,b)}\pi_1$\,.
The subscripts $2r$ $(2\bar{r})$ denotes the left- (right-) moving Ramond number defined by
\begin{center}
 Ramond No. $=$ No. of Ramond inputs $-$ No. of Ramond outputs.
\end{center}
Although it is not difficult to construct the products $\bd{B}$ to be cyclic
(with respect to $\omega_l$), they cannot be used as $\bd{L}$ in the action (\ref{action}) 
since they have wrong picture numbers except for the first component (\ref{B11}). 
The picture number deficit of the components (\ref{B21}), 
(\ref{B12}), and (\ref{B22}) from those of $\bd{L}$ are $[1,0]$\,, $[0,1]$ and $[1,1]$\,,
respectively: the picture number of their outputs are $(-3/2,-1)$\,, $(-1,-2/3)$\,, 
and $(-3/2,-3/2)$\,, respectively.

Using the coderivation $\bd{B}$\,, we can construct three independent $L_\infty$ algebras,
$(\mathcal{H}_l, \bd{D})$\,, $(\mathcal{H}_l, \bd{C})$ and $(\mathcal{H}_l, \bar{\bd{C}})$\,, 
which are called an $L_\infty$ triplet in \cite{Matsunaga:2016zsu}, with
\begin{subequations} 
\begin{align}
 \pi_1\bd{D}\ =&\ \pi_1\bd{Q} + \pi_1^{(0,0)}\bd{B}\,,\\
 \pi_1\bd{C}\ =&\ \pi_1\bd{\eta} - \pi_1^{(1,0)}\bd{B} - \frac{1}{2}\bar{X}\pi_1^{(1,1)}\bd{B}\,,\\
 \pi_1\bar{\bd{C}}\ =&\ \pi_1\bar{\bd{\eta}} - \pi_1^{(0,1)}\bd{B} - \frac{1}{2}X\pi_1^{(1,1)}\bd{B}\,.
\end{align}
\end{subequations}
We can merge them into an $L_\infty$ algebra, $(\mathcal{H}_l,\bd{D}-\bd{C}-\bar{\bd{C}})$\,, as
\begin{equation}
 \pi_1(\bd{D}-\bd{C}-\bar{\bd{C}})\ =\ \pi_1\bd{Q}-\pi_1\bd{\eta}-\pi_1\bar{\bd{\eta}}
%+ (\pi_1^{(0,0)}+\pi_1^{(1,0)}+\pi_1^{(0,1)})\bd{B} 
+ (\pi_1-\pi_1^{(1,1)})\bd{B} 
%+ (1-\pi^{(1,1)})\pi_1\bd{B} 
+ \frac{1}{2}(X+\bar{X})\pi_1^{(1,1)}\bd{B}\,,
\end{equation}
because the $L_\infty$ relation 
\begin{equation}
 [\,\bd{D}-\bd{C}-\bar{\bd{C}}\,, \bd{D}-\bd{C}-\bar{\bd{C}}\,]\ =\ 0\,,
\end{equation}
holds for each picture number deficit and can decompose as 
\begin{align}\label{L infinity}
\begin{split} 
&\ [\,\bd{D}\,, \bd{D}\,]\ =\ [\,\bd{C}\,, \bd{C}\,]\ 
=\ [\,\bar{\bd{C}}\,, \bar{\bd{C}}\,]\ =\ 0\,,\\
&\ [\,\bd{D}\,, \bd{C}\,]\ =\ [\,\bd{D}\,, \bar{\bd{C}}\,]\ 
=\ [\,\bd{C}\,, \bar{\bd{C}}\,]\ =\ 0\,.
\end{split}
\end{align}
Note that it is difficult to make this merged $L_\infty$ algebra
$(\mathcal{H}_l,\bd{D}-\bd{C}-\bar{\bd{C}})$ cyclic, unlike the heterotic string field theory case,
due to the last (R-R) component.
Once this $L_\infty$ triplet is constructed, we can transform it into the triplet 
$(\bd{\eta},\bar{\bd{\eta}};\bd{L})$ closed in $\mathcal{H}^{res}$ as
\begin{subequations} 
\begin{align}
&\ \pi_1\hat{\bd{F}}^{-1}\bd{C}\hat{\bd{F}}\ =\ \pi_1\bd{\eta}\,,\qquad
 \pi_1\hat{\bd{F}}^{-1}\bar{\bd{C}}\hat{\bd{F}}\ =\ \pi_1\bar{\bd{\eta}}\,,\\
\pi_1\bd{L}\, \equiv&\ \pi_1\hat{\bd{F}}^{-1}\bd{D}\hat{\bd{F}} = \pi_1\bd{Q} 
+ \mathcal{G}\pi_1\bd{b}\,,\qquad
\pi_1\bd{b}\ =\ \pi_1\bd{B}\hat{\bd{F}}\,.
%+ \mathcal{G}\pi_1\bd{b}\,,
%\qquad \pi_1\bd{b}\ =\ \pi_1\bd{B}\hat{\bd{F}}\,,
\label{def L}
\end{align}
\end{subequations}
by using the cohomomorphism
\begin{equation}
 \pi_1\hat{\bd{F}}^{-1}\ =\ \pi_1\id-\mathfrak{S}\pi_1\bd{B}\,,\qquad
 \mathfrak{S}\ =\
\Xi\pi^{(1,0)}+\bar{\Xi}\pi^{(0,1)} 
+ \frac{1}{2}(\Xi\bar{X}+X\bar{\Xi})\pi^{(1,1)}\,.
\label{def F}
\end{equation}
%\begin{equation}
% \pi_1\hat{\bd{F}}^{-1}\ =\ \pi_1\id -\Xi \pi_1^{(1,0)}\bd{B}
%- \bar{\Xi}\pi_1^{(0,1)}\bd{B}-\frac{1}{2}(\Xi\bar{X}+X\bar{\Xi})\pi_1^{(1,1)}\bd{B}\,.
%\end{equation}
As was shown in \cite{Kunitomo:2019glq}, if $\bd{B}$ is cyclic with respect 
to $\omega_l$\,, $\bd{L}$ is cyclic with respect to 
$\Omega$\,, and thus gives a cyclic $L_\infty$ algebra 
$(\mathcal{H}^{res},\Omega,\bd{L})$ %\,, which we use for constructing the action.
used for the action.

\subsection{Explicit construction}\label{construction}

Now, the task is to construct odd coderivation $\bd{B}$ satisfying the $L_\infty$
relations (\ref{L infinity}) written for $\bd{B}$ as
\begin{subequations}\label{L infinity B}  
\begin{align}
[\,\bd{Q}\,,\bd{B}\,] + \frac{1}{2}[\,\bd{B}\,, \bd{B}\,]^{11}\ =&\ 0\,,\\
[\,\bd{\eta}\,, \bd{B}\,] - \frac{1}{2}[\,\bd{B}\,, \bd{B}\,]^{21}
-\frac{1}{4}[\,\bd{B}\,, \bd{B}\,]^{22}_{\bar{X}}\ =&\ 0\,,\\
[\,\bar{\bd{\eta}}\,, \bd{B}\,] - \frac{1}{2}[\,\bd{B}\,, \bd{B}\,]^{12}
-\frac{1}{4}[\,\bd{B}\,, \bd{B}\,]^{22}_X\ =&\ 0\,.
\end{align}
\end{subequations}
Here, the square bracket $[\cdot,\cdot]^{ab}$ with $a,b=1$ or $2$ denotes the projected commutator 
defined by projecting onto specific cyclic Ramond numbers, an extension of that introduced 
in \cite{Kunitomo:2019glq} for the heterotic string field theory.
In addition, similar bracket with subscript $\mathcal{X}=X$ or $\bar{X}$ is defined by 
inserting $\mathcal{X}$ at the intermediate state\footnote{
This bracket cannot always be well-defined since $X$ and $\bar{X}$ are the PCO's acting 
on the states with picture number $-3/2$\,.}. 
We give an explicit definition of these brackets in Appendix~\ref{pro comm}, 
in which we also summarize the Jacobi identities they satisfy.

Assume that the cyclic $L_\infty$ algebra 
$(\mathcal{H}_l,\omega_l,\bd{L}^{(0,0)}=\sum_{n=0}^\infty\bd{L}^{(0,0)}_{n+1})$ 
without picture number, which is straightforwardly constructed similar to that 
of the bosonic string field theory \cite{Saadi:1989tb,Kugo:1989aa,Zwiebach:1992ie},
is known\footnote{
%As shown shortly,
This is actually a cyclic $L_\infty$ algebra $(\mathcal{H},\omega_s,\bd{L}^{(0,0)})$
closed in the small Hilbert space.
}. 
We define a generating functional
\begin{equation}
 \bd{L}^{(0,0)}(s,\bar{s})\ =\ \bd{Q} + \sum_{m,\bar{m},r,\bar{r}}s^m\bar{s}^{\bar{m}}
\bd{L}^{(0,0)}_{m+r+1,\bar{m}+\bar{r}+1}\mid^{(2r,2\bar{r})}\
\equiv\ \bd{Q} + \bd{L}^B(s,\bar{s})\,,
\end{equation}
with
\begin{equation}
 \qquad
\bd{L}^{(0,0)}_{m+r+1,\bar{m}+\bar{r}+1}\mid^{(2r,2\bar{r})}\ =\
\delta_{m+r,\bar{m}+\bar{r}}
\bd{L}^{(0,0)}_{m+r+1}\mid^{(2r,2\bar{r})}\,,
\end{equation}
which reduces to $\bd{L}^{(0,0)}$ at $(s,\bar{s})=(1,1)$\,.
The parameter $s$ or $\bar{s}$ is counting the left- or right-moving picture number 
deficit from $\bd{B}$\,,
%$\bd{B}^{(p,\bar{p})}_{p+r+1,\bar{p}+\bar{r}+1}\mid^{(2r,2\bar{r})}$\,,
respectively. We can show that $\bd{L}^B(s,\bar{s})$ satisfies
\begin{subequations}\label{L infinity 0} 
 \begin{align}
 [\,\bd{Q}\,, \bd{L}^B(s,\bar{s})\,] 
&\ 
+ \frac{1}{2}[\,\bd{L}^B(s,\bar{s})\,, \bd{L}^B(s,\bar{s})\,]^{11}
+ \frac{s}{2}[\,\bd{L}^B(s,\bar{s})\,, \bd{L}^B(s,\bar{s})\,]^{21}
\nonumber\\
&\ 
+ \frac{\bar{s}}{2}[\,\bd{L}^B(s,\bar{s})\,, \bd{L}^B(s,\bar{s})\,]^{12}
+ \frac{s\bar{s}}{2}[\,\bd{L}^B(s,\bar{s})\,, \bd{L}^B(s,\bar{s})\,]^{22}\
=\ 0\,,
\label{L infinity bosonic}
\end{align}
derived from the $L_\infty$ relation $[\bd{L}^{(0,0)},\bd{L}^{(0,0)}]=0$\,.
It is also closed in the small Hilbert space,
 \begin{equation}
%%%%%%%%%%%%
 [\,\bd{\eta}\,, \bd{L}^B(s,\bar{s})\,]\ \ =\ 0\,,\qquad
%%%%%%%%%%%%
 [\,\bar{\bd{\eta}}\,, \bd{L}^B(s,\bar{s})\,]\ \ =\ 0\,,
\label{small bosonic}
 \end{equation}
\end{subequations}
since it can be constructed without using $\xi$ or $\bar{\xi}$.
%Then, we extend $\bd{B}$
%$\bd{B}^{(p,\bar{p})}_{p+r+1,\bar{p}+\bar{r}+1}\mid^{(2r,2\bar{r})}$
%to $\bd{B}^{(p,\bar{p})}_{p+m+r+1,\bar{p}+\bar{m}+\bar{r}+1}\mid^{(2r,2\bar{r})}$
%that has picture number deficit from $\bd{B}^{(p,\bar{p})}_{p+r+1,\bar{p}+\bar{r}+1}\mid^{(2r,2\bar{r})}$
%by $[m,\bar{m}]$
Then we extend $\bd{B}$
to include those with non-zero picture number deficit from $\bd{B}$ and define a generating functional 
\begin{align}
\begin{split}
  \bd{B}\sst\ =&\ \sum_{p,m,r=0}^\infty\sum_{\bar{p},\bar{r},\bar{m}=0}^\infty
t^{p+\bar{p}}s^m\bar{s}^{\bar{m}}\bd{B}^{(p,\bar{p})}_{p+m+r+1,\bar{p}+\bar{m}+\bar{r}+1}\mid^{(2r,2\bar{r})}\\
\equiv&\ \sum_{p,\bar{p}=0}^\infty t^{p+\bar{p}}\bd{B}^{(p,\bar{p})}(s,\bar{s})\,,
%=\ \sum_{m,\bar{m}=0}^\infty s^m \bar{s}^{\bar{m}}\bd{B}^{[m,\bar{m}]}(t)\,,
\end{split}
\end{align}
by introducing another parameter $t$ counting the (total) picture number. 
The $L_\infty$ relations  (\ref{L infinity B}) are extended for $\bd{B}\sst$ to
\begin{subequations} \label{ext L infinity}
\begin{align}
\bd{I}\sst\ \equiv&\ 
[\bd{Q}\,,\bd{B}(s,\bar{s},t)] + \frac{1}{2}[\bd{B}(s,\bar{s},t),\bd{B}(s,\bar{s},t)]^{11}\ 
\nonumber\\
&\ + \frac{s}{2} \Big([\bd{B}(s,\bar{s},t)\,, \bd{B}(s,\bar{s},t)]^{21}
+ t[\bd{B}(s,\bar{s},t)\,, \bd{B}(s,\bar{s},t)]^{22}_{\bar{X}}\Big)
\nonumber\\
&\ + \frac{\bar{s}}{2} \Big([\bd{B}(s,\bar{s},t)\,, \bd{B}(s,\bar{s},t)]^{12}
+ t[\bd{B}(s,\bar{s},t)\,, \bd{B}(s,\bar{s},t)]^{22}_X\Big)
\nonumber\\
&\ + \frac{s\bar{s}}{2}[\bd{B}(s,\bar{s},t)\,, \bd{B}(s,\bar{s},t)]^{22}\ =\ 0\,,
\label{L infinity I}\\
%%%%%%%%%%%%%%%%%%%%%%%
\bd{J}(s,\bar{s},t)\ \equiv&\ [\bd{\eta}\,,\bd{B}(s,\bar{s},t)] 
\nonumber\\
&\
- \frac{t}{2}\left([\bd{B}(s,\bar{s},t)\,, \bd{B}(s,\bar{s},t)]^{21}
+ \frac{t}{2}[\bd{B}(s,\bar{s},t)\,, \bd{B}(s,\bar{s},t)]^{22}_{\bar{X}}\right)\ =\ 0\,,
\label{L infinity J}\\
%%%%%%%%%%%%%%%%%%%%%%%
\bar{\bd{J}}(s,\bar{s},t)\ \equiv&\ [\bar{\eta}, \bd{B}(s,\bar{s},t)]
\nonumber\\
&\
- \frac{t}{2}\left([\bd{B}(s,\bar{s},t)\,, \bd{B}(s,\bar{s},t)]^{12}
+ \frac{t}{2}[\bd{B}(s,\bar{s},t)\,, \bd{B}(s,\bar{s},t)]^{22}_X\right)\ =\ 0\,.
\label{L infinity Jbar}
\end{align}
\end{subequations}
The string product $\bd{B}$ is obtained as $\bd{B}=\bd{B}(0,0,1)$\,,
and the relations (\ref{ext L infinity}) reduce to (\ref{L infinity B})
for $\bd{B}$\,.
%Then let us show that the relations (\ref{ext L infinity}) hold 
%if $\bd{B}\sst$ satisfies the differntial equations
We can show that if $\bd{B}\sst$ satisfies the differential equations,
\begin{subequations}\label{diff eqs} 
 \begin{align}
 \partial_t\bd{B}(s,\bar{s},t)\ =&\ 
[\bd{Q}\,, \llb\sst] + [\bd{B}\sst\,, \llb\sst]^{11}
\nonumber\\
&\
+ s\Big([\bd{B}\sst\,,\llb\sst]^{21}+t[\bd{B}\sst\,, \llb\sst]^{22}_{\bar{X}}\Big)
\nonumber\\
&\
+ \bar{s}\Big([\bd{B}\sst\,, \llb\sst]^{12}+t[\bd{B}\sst\,, \llb\sst]^{22}_{X}\Big)
\nonumber\\
&\
+ s\bar{s}[\bd{B}\sst\,, \llb\sst]^{22}
\nonumber\\
&\
+\frac{s}{2}[\bd{B}(s,\bar{s},t)\,, \bd{B}(s,\bar{s},t)]^{22}_{\bar{\Xi}}
+\frac{\bar{s}}{2}[\bd{B}(s,\bar{s},t)\,, \bd{B}(s,\bar{s},t)]^{22}_{\Xi}\,,
\label{diff eq t}\\
%%%%%%%%%%%%%%%%%%
\partial_s\bd{B}(s,\bar{s},t)\ 
=&\ 
[\bd{\eta}\,,\bd{\lambda}(s,\bar{s},t)]\ 
\nonumber\\
&\
- t\Big([\bd{B}\sst\,, \llb\sst]^{21}+\frac{t}{2}[\bd{B}\sst\,, \llb\sst]^{22}_{\bar{X}}\Big)\,,
\label{diff eq s}\\
%+ t \left([\bd{B}(s,\bar{s},t)\,, \bd{\lambda}(s,\bar{s},t)+\bar{\bd{\lambda}}(s,\bar{s},t)]^{21}
%+ \frac{t}{2}[\bd{B}(s,\bar{s},t)\,, 
%\bd{\lambda}(s,\bar{s},t)+\bar{\bd{\lambda}}(s,\bar{s},t)]^{22}_{\bar{X}}\right)\,,\\
%%%%%%%%%%%%%%%%%%
\partial_{\bar{s}}\bd{B}(s,\bar{s},t)\ =&\ 
[\bar{\bd{\eta}}\,,\bar{\bd{\lambda}}(s,\bar{s},t)] 
\nonumber\\
&\
- t\Big([\bd{B}\sst\,, \llb\sst]^{12} + \frac{t}{2}[\bd{B}\sst\,, \llb\sst]^{22}_X\Big)\,,
\label{diff eq sbar}
%+ t \left([\bd{B}(s,\bar{s},t)\,, \bd{\lambda}(s,\bar{s},t)+\bar{\bd{\lambda}}(s,\bar{s},t)]^{12}
%+ \frac{t}{2}[\bd{B}(s,\bar{s},t)\,, \bd{\lambda}(s,\bar{s},t)+\bar{\bd{\lambda}}(s,\bar{s},t)]^{22}_{X}\right)\,,
\end{align}
\end{subequations}
the $L_\infty$ relations (\ref{ext L infinity}) hold. 
Here, we introduced two degree-even coderivations $\bd{\lambda}\sst$ and $\bar{\bd{\lambda}}\sst$
satisfying 
\begin{equation}
 [\bd{\eta}\,, \bar{\bd{\lambda}}(s,\bar{s},t)]\ =\ 0\,,\qquad
 [\bar{\bd{\eta}}\,, \bd{\lambda}(s,\bar{s},t)]\ =\ 0\,,
\end{equation}
and used an abbreviated notation
\begin{equation}
 \bd{\lambda}(s,\bar{s},t)+\bar{\bd{\lambda}}(s,\bar{s},t)\
=\ \llb\sst\,.
\end{equation}
These degree-even coderivations, $\bd{\lambda}\sst$ and $\bar{\bd{\lambda}}\sst$\,, 
are called (generating functionals of) gauge products and can be expanded in the parameters as
\begin{align}
 \bd{\lambda}\sst\ =&\ \sum_{p,m,r=0}^\infty\sum_{\bar{p},\bar{r},\bar{m}=0}^\infty
t^{p+\bar{p}}s^m\bar{s}^{\bar{m}}\bd{\lambda}^{(p+1,\bar{p})}_{p+m+r+2,\bar{p}+\bar{m}+\bar{r}+1}\mid^{(2r,2\bar{r})}
\nonumber\\
\equiv&\ \sum_{p,\bar{p}=0}^\infty t^{p+\bar{p}} \bd{\lambda}^{(p+1,\bar{p})}(s,\bar{s})\,,\\
%=\ \sum_{m,\bar{m}}^\infty s^m\bar{s}^{\bar{m}}\bd{\lambda}^{[m,\bar{m}]}(t)\,,\\
%%%%%%%%%%%%
\bar{\bd{\lambda}}\sst\ =&\ \sum_{p,m,r=0}^\infty\sum_{\bar{p},\bar{r},\bar{m}=0}^\infty
t^{p+\bar{p}}s^m\bar{s}^{\bar{m}}
\bar{\bd{\lambda}}^{(p,\bar{p}+1)}_{p+m+r+1,\bar{p}+\bar{m}+\bar{r}+2}\mid^{(2r,2\bar{r})}
\nonumber\\
\equiv&\ \sum_{p,\bar{p}=0}^\infty t^{p+\bar{p}} \bar{\bd{\lambda}}^{(p,\bar{p}+1)}(s,\bar{s})\,.
%=\ \sum_{m,\bar{m}}^\infty s^m\bar{s}^{\bar{m}}\bar{\bd{\lambda}}^{[m,\bar{m}]}(t)\,.
\end{align}
The bracket with subscript, $[\cdot,\cdot]^{22}_{\mathfrak{X}}$ with
$\mathfrak{X}=\Xi$ or $\bar{\Xi}$\,, is the (graded) commutator with
$\mathfrak{X}$ inserted at the intermediate state, whose explicit definition
is given in Appendix~\ref{pro comm}.
By differentiating (\ref{L infinity I}) by $t$ and
using the differential equation (\ref{diff eq t}), we find that
\begin{align}
 \partial_t\bd{I}(s,\bar{s},t)\ =&\ 
[\bd{I}\sst\,, \llb\sst]^{11}
\nonumber\\
&\
+ s\Big([\bd{I}\sst\,, \llb\sst]^{21}+t[\bd{I}\sst\,, \llb\sst]^{22}_{\bar{X}}\Big)
\nonumber\\
&\
+ \bar{s}\Big([\bd{I}\sst\,, \llb\sst]^{12}+t[\bd{I}\sst\,, \llb\sst]^{22}_X
\Big)
\nonumber\\
&\
+ s\bar{s}[\bd{I}\sst\,, \llb\sst]^{22}
\nonumber\\
&\
+ s[\bd{I}(s,\bar{s},t)\,, \bd{B}(s,\bar{s},t)]^{22}_{\bar{\Xi}}
+ \bar{s}[\bd{I}(s,\bar{s},t)\,, \bd{B}(s,\bar{s},t)]^{22}_{\Xi}\,,
\end{align}
which implies if $\bd{I}(s,\bar{s},0)=0$\,, then $\bd{I}\sst=0$ for $^\forall t$\,.
Similarly, by differentiating (\ref{L infinity J}) and (\ref{L infinity Jbar}) by $t$
and using the differential equations (\ref{diff eqs}) we obtain
\begin{subequations}
\begin{align}
%%%%%%%%%%%%%%%%
\partial_t\bd{J}(s,\bar{s},t)\ =&\ -\partial_s\bd{I}(s,\bar{s},t) 
+ [\bd{J}\sst\,, \llb\sst]^{11}
\nonumber\\
&\
+ s\Big([\bd{J}\sst\,, \llb\sst]^{21}+t[\bd{J}\sst\,, \llb\sst]^{22}_{\bar{X}}\Big)
\nonumber\\
&\
+ \bar{s}\Big([\bd{J}\sst\,, \llb\sst]^{12}+t[\bd{J}\sst\,, \llb\sst]^{22}_X\Big)
\nonumber\\
&\
+ s\bar{s}[\bd{J}\sst\,, \llb\sst]^{22}
\nonumber\\
&\
- t \Big([\bd{I}\sst\,, \llb\sst]^{21} + \frac{t}{2}[\bd{I}\sst\,, \llb\sst]^{22}_{\bar{X}}
\Big)
\nonumber\\
&\
+ s[\bd{J}(s,\bar{s},t)\,, \bd{B}(s,\bar{s},t)]^{22}_{\bar{\Xi}}
+ \bar{s}[\bd{J}(s,\bar{s},t)\,, \bd{B}(s,\bar{s},t)]^{22}_{\Xi}\,,\\
%%%%%%%%%%%%%%%%%
\partial_t\bar{\bd{J}}(s,\bar{s},t)\ =&\ -\partial_{\bar{s}}\bd{I}(s,\bar{s},t) 
+ [\bar{\bd{J}}\sst\,, \llb\sst]^{11}
\nonumber\\
&\
+ s \Big([\bar{\bd{J}}\sst\,, \llb\sst]^{21} + t[\bar{\bd{J}}\sst\,, \llb\sst]^{22}_{\bar{X}}\Big)
\nonumber\\
&\
+ \bar{s}\Big([\bar{\bd{J}}\sst\,, \llb\sst]^{12} + t[\bar{\bd{J}}\sst\,, \llb\sst]^{22}_X
\Big)
\nonumber\\
&\
+ s\bar{s}[\bar{\bd{J}}\sst\,, \llb\sst]^{22}
\nonumber\\
&\
-t \Big([\bd{I}\sst\,, \llb\sst]^{12} + \frac{t}{2}[\bar{\bd{J}}\sst\,, \llb\sst]^{22}_X\Big)
\nonumber\\
&\
+ s[\bar{\bd{J}}(s,\bar{s},t)\,, \bd{B}(s,\bar{s},t)]^{22}_{\bar{\Xi}}
+ \bar{s}[\bar{\bd{J}}(s,\bar{s},t)\,, \bd{B}(s,\bar{s},t)]^{22}_{\Xi}\,,
\end{align}
\end{subequations}
which imply if $\bd{I}\sst=0$ and $\bd{J}(s,\bar{s},0)=\bar{\bd{J}}(s,\bar{s},0)=0$\,, then
$\bd{J}\sst=\bar{\bd{J}}\sst=0$ for $^\forall t$\,.
On the other hand, (\ref{ext L infinity}) reduce to
\begin{subequations} 
 \begin{align}
 \bd{I}(s,\bar{s},0)\ =&\ [\bd{Q}\,,\bd{B}(s,\bar{s},0)] 
+ \frac{1}{2}[\bd{B}(s,\bar{s},0),\bd{B}(s,\bar{s},0)]^{11}
+ \frac{s}{2} [\bd{B}(s,\bar{s},0)\,, \bd{B}(s,\bar{s},0)]^{21}
\nonumber\\
&\hspace{18mm}
+ \frac{\bar{s}}{2} [\bd{B}(s,\bar{s},0)\,, \bd{B}(s,\bar{s},0)]^{12}
+ \frac{s\bar{s}}{2}[\bd{B}(s,\bar{s},0)\,, \bd{B}(s,\bar{s},0)]^{22}\ =\ 0\,,
\end{align}
\begin{equation}
 \bd{J}(s,\bar{s},0)\ =\ [\bd{\eta}\,, \bd{B}(s,\bar{s},0)]\ =\ 0\,,\qquad
%%%%%%%%%%%%%
\bar{\bd{J}}(s,\bar{s},0)\ =\ [\bar{\bd{\eta}}\,, \bd{B}(s,\bar{s},0)]\ =\ 0\,,
 \end{equation}
\end{subequations}
at $t=0$\,, %which are satisfied by setting 
which hold if we set
\begin{equation}
\bd{B}(s,\bar{s},0) =\ \bd{B}^{(0,0)}(s,\bar{s})\ =\ \bd{L}^B(s,\bar{s})
\label{initial}
\end{equation}
as was seen in (\ref{L infinity 0}).
Thus, if $\bd{B}\sst$ satisfies the differential equations (\ref{diff eqs})
with the initial condition (\ref{initial}),
the relations (\ref{ext L infinity}) hold. We obtain the string product $\bd{B}$
satisfying the relations (\ref{construction}) as $\bd{B}=\bd{B}(0,0,1)$\,. 
 
\begin{figure}[hbtp]
 \begin{center}
  \includegraphics[clip,width=10cm]{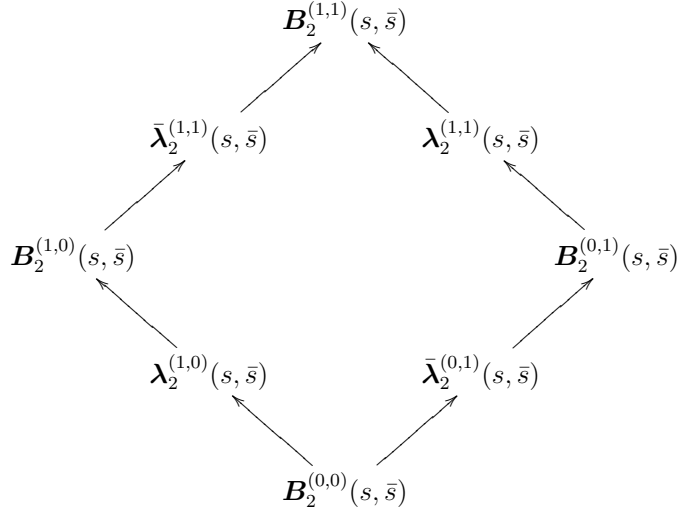}
    \caption{The flow of how the 2-string (gauge) products are determined.}
      \label{2string ss}
 \end{center}
\end{figure}

\begin{figure}[hbtp]
  \begin{center}
   \includegraphics[clip,width=15cm]{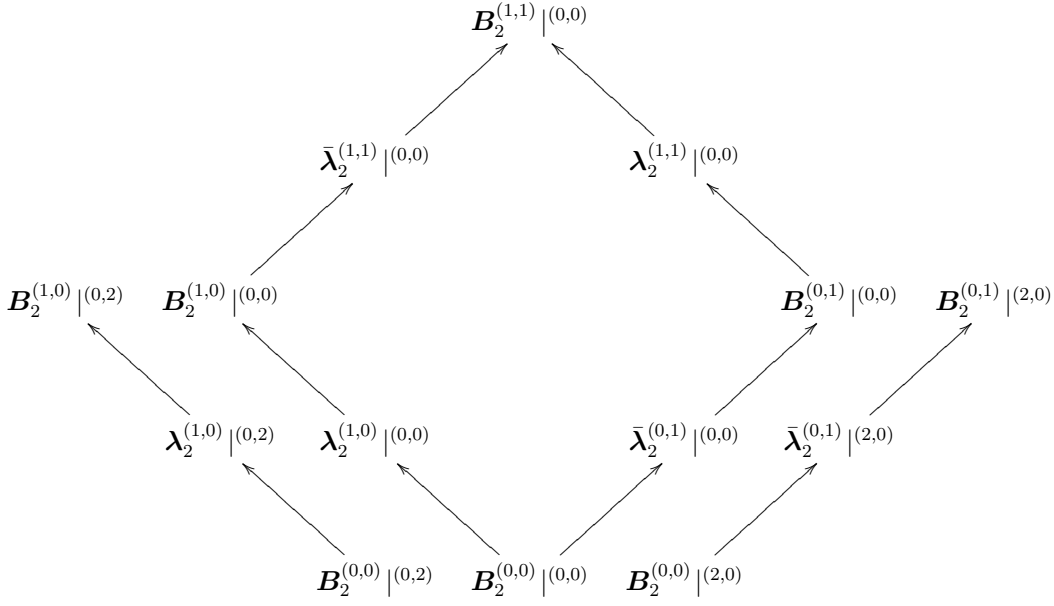}
\caption{The flow of how each component of the 2-string (gauge) products is determined.}
\label{2string comp}
  \end{center}
\end{figure}

Under the initial condition (\ref{initial}), 
we can explicitly solve the differential equations (\ref{diff eqs}) 
and find $\bd{B}\sst=\sum_{n=2}^\infty\bd{B}_n\sst$ in ascending order of $n$.
First, the 2-string products $\bd{B}_2\sst$\,,
$\bd{\lambda}_2\sst$\,, and $\bar{\bd{\lambda}}_2\sst$ are expanded in $t$ as
\begin{subequations} 
\begin{align}
 \bd{B}_2(s,\bar{s},t)\ =&\ \bd{B}_2^{(0,0)}(s,\bar{s}) + 
t \left(\bd{B}_2^{(1,0)}(s,\bar{s}) + \bd{B}_2^{(0,1)}(s,\bar{s})\right) +
t^2 \bd{B}_2^{(1,1)}(s,\bar{s})\,,\\
%%%%%%%%%%%%%%
\bd{\lambda}_2(s,\bar{s},t)\ =&\ \bd{\lambda}_2^{(1,0)}(s,\bar{s}) +
t \bd{\lambda}_2^{(1,1)}(s,\bar{s})\,,\\
%%%%%%%%%%%%%%
\bar{\bd{\lambda}}_2(s,\bar{s},t)\ =&\ \bar{\bd{\lambda}}_2^{(0,1)}(s,\bar{s}) +
t \bar{\bd{\lambda}}_2^{(1,1)}(s,\bar{s})\,.
\end{align}
\end{subequations}
We can decompose the equations (\ref{diff eqs}) to
\begin{subequations}\label{diff 2point}
 \begin{alignat}{2}
\bd{B}_2^{(1,0)}(s,\bar{s})\ =&\ [\,\bd{Q}\,, \bd{\lambda}^{(1,0)}(s,\bar{s})\,]\,,\qquad&
\bd{B}_2^{(0,1)}(s,\bar{s})\ =&\ [\,\bd{Q}\,, \bar{\bd{\lambda}}^{(1,0)}(s,\bar{s})\,]\,,
\label{2point t}\\
%%%%%%%%%%%%%%%
\partial_s\bd{B}_2^{(0,0)}(s,\bar{s})\ =&\ [\,\bd{\eta}\,, \bd{\lambda}_2^{(1,0)}(s,\bar{s})\,]\,,\qquad&
\partial_{\bar{s}}\bd{B}_2^{(0,0)}(s,\bar{s})\ =&\ [\,\bar{\eta}\,, \bar{\bd{\lambda}}_2^{(0,1)}(s,\bar{s})\,]\,.
\label{2point s sbar}\\
%%%%%%%%%%%%%%%
2 \bd{B}_2^{(1,1)}(s,\bar{s})\ =&\ [\,\bd{Q}\,, \bd{\lambda}_2^{(1,1)}(s,\bar{s}) 
+ \bar{\bd{\lambda}}_2^{(1,1)}(s,\bar{s})\,]\,,& &
\label{2point t 2}\\
%%%%%%%%%%%%%%%
\partial_s\bd{B}_2^{(0,1)}(s,\bar{s})\ =&\ [\,\bd{\eta}\,, \bd{\lambda}_2^{(1,1)}(s,\bar{s})\,]\,,\qquad&
\partial_{\bar{s}}\bd{B}_2^{(1,0)}(s,\bar{s})\ =&\ [\,\bar{\eta}\,, \bar{\bd{\lambda}}_2^{(1,1)}(s,\bar{s})\,]\,,
\label{2point s sbar 2}
\end{alignat}
\end{subequations}
for each order of $t$ and each picture number.
Note that there is no nonlinear term in these equations (\ref{diff 2point}) for the 2-string products.
First,
eqs.~(\ref{2point s sbar}) can be solved for $\bd{\lambda}_2^{(1,0)}(s,\bar{s})$ 
and $\bar{\bd{\lambda}}_2^{(0,1)}(s,\bar{s})$ as
\begin{subequations}\label{sol 1} 
\begin{align}
\pi_1\bd{\lambda}_2^{(1,0)}(s,\bar{s})\ =&\ 
\frac{1}{3}\left(\xi \partial_sL^B_2(s,\bar{s}) 
- \partial_sL^B_2(s,\bar{s})(\xi\pi_1\wedge\id)\right)\ \equiv\ 
\xi\circ \pi_1\partial_s\bd{L}^B_2(s,\bar{s})\,,\\
%%%%%%%%%
\pi_1\bar{\bd{\lambda}}_2^{(0,1)}(s,\bar{s})\ =&\  
\frac{1}{3}\left(\bar{\xi} \partial_{\bar{s}}L^B_2(s,\bar{s}) 
- \partial_{\bar{s}}L^B_2(s,\bar{s})(\bar{\xi}\pi_1\wedge\id)\right)\ \equiv\
\bar{\xi}\circ\pi_1\partial_{\bar{s}}\bd{L}^B_2(s,\bar{s})\,,
 \end{align}
\end{subequations}
under the initial condition $\bd{B}_2^{(0,0)}(s,\bar{s})=\bd{L}_2^B(s,\bar{s})$\,.
Then, eqs. (\ref{2point t}) determine $\bd{B}_2^{(1,0)}(s,\bar{s})$ 
and $\bd{B}_2^{(0,1)}(s,\bar{s})$ as
\begin{subequations}\label{sol 2} 
\begin{align}
 \pi_1\bd{B}_2^{(1,0)}(s,\bar{s})\ =&\ \frac{1}{3}\left(
X_0\partial_s L^B_2(s,\bar{s}) + \partial_s L^B_2(s,\bar{s})(X_0\pi\wedge\id_2)\right)\
\equiv\ X\circ\pi_1\partial_s\bd{L}_2^B(s,\bar{s})\,,\\
 \pi_1\bd{B}_2^{(0,1)}(s,\bar{s})\ =&\ \frac{1}{3}\left(
\bar{X}_0\partial_s L^B_2(s,\bar{s}) + \partial_s L^B_2(s,\bar{s})(\bar{X}_0\pi\wedge\id_2)\right)\
\equiv\ \bar{X}\circ\pi_1\partial_{\bar{s}}\bd{L}_2^B(s,\bar{s})\,.
\end{align}
\end{subequations}
Using these results, the equations (\ref{2point s sbar 2}) are solved for $\pi_1\bd{\lambda}^{(1,1)}(s,\bar{s})$
and $\pi_1\bar{\bd{\lambda}}^{(1,1)}(s,\bar{s})$  as
\begin{subequations}\label{sol 3} 
 \begin{align}
\pi_1\bd{\lambda}^{(1,1)}(s,\bar{s})\ =&\ \xi\circ\pi_1\partial_s\bd{B}_2^{(0,1)}(s,\bar{s})\
%=\ \xi\circ\partial_s\left(\bar{X}\circ\pi_1\partial_{\bar{s}}\bd{L}_2^B(s,\bar{s})\right)\,,\\
= \xi\circ\bar{X}\circ\pi_1\partial_s\partial_{\bar{s}}\bd{L}_2^B(s,\bar{s})\,,\\
\pi_1\bar{\bd{\lambda}}^{(1,1)}(s,\bar{s})\ =&\ \bar{\xi}\circ\pi_1\partial_{\bar{s}}\bd{B}_2^{(1,0)}(s,\bar{s})\
%=\ \bar{\xi}\circ\partial_{\bar{s}}\left(X\circ\pi_1\partial_s\bd{L}_2^B(s,\bar{s})\right)\,,
= \bar{\xi}\circ X\circ\pi_1\partial_s\partial_{\bar{s}}\bd{L}_2^B(s,\bar{s})\,,
\end{align}
\end{subequations}
which determine $\bd{B}_2^{(1,1)}(s,\bar{s})$ as
\begin{equation}
 \pi_1\bd{B}_2^{(1,1)}(s,\bar{s})\ 
%=&\  \frac{1}{2}\left(
%X\circ\pi_1\partial_s\bd{B}_2^{(0,1)}(s,\bar{s}) 
%+ \bar{X}\circ\pi_1\partial_{\bar{s}}\bd{B}_2^{(1,0)}(s,\bar{s})\right)
%\nonumber\\
%=&\ \frac{1}{2}\left(
%X\circ\partial_s\left(\bar{X}\circ\pi_1\partial_{\bar{s}} \bd{L}_2^{B}(s,\bar{s})\right) 
%+ \bar{X}\circ\partial_{\bar{s}}\left(X\circ\pi_1\partial_s \bd{L}_2^{B}(s,\bar{s})\right)\right)\,,
=\ X\circ\bar{X}\circ\pi_1\partial_s\partial_{\bar{s}}\bd{L}_2^B(s,\bar{s})\,,
\label{sol 4}
\end{equation}
from (\ref{2point t 2}). We illustrate in Fig.~\ref{2string ss} the flow of how the 2-string (gauge) 
products are determined.
The string and gauge products with specific cyclic Ramond numbers are obtained
as coefficients by expanding in $s$ and $\bar{s}$\,:
 \begin{align}
% \bd{L}_2^{B}(s,\bar{s})\ =&\ \bd{L}_2^{B}\mid^{(2,2)} + s\bd{L}_2^{B}\mid^{(0,2)}
%+ \bar{s}\bd{L}_2^{B}\mid^{(2,0)} + s\bar{s}\bd{L}_2^{B}\mid^{(0,0)}\,,\\
%%%%%%%%%%%%%%%
 \bd{B}_2^{(0,0)}(s,\bar{s})\ =&\ \bd{B}_2^{(0,0)}\mid^{(2,2)} + s\bd{B}_2^{(0,0)}\mid^{(0,2)}
+ \bar{s}\bd{B}_2^{(0,0)}\mid^{(2,0)} + s\bar{s}\bd{B}_2^{(0,0)}\mid^{(0,0)}\,,\\
%%%%%%%%%%%%%
 \bd{B}_2^{(1,0)}(s,\bar{s})\ =&\ \bd{B}_2^{(1,0)}\mid^{(0,2)} + \bar{s}\bd{B}_2^{(1,0)}\mid^{(0,0)}\,,\\
%%%%%%%%%%%%%
 \bd{B}_2^{(0,1)}(s,\bar{s})\ =&\ \bd{B}_2^{(0,1)}\mid^{(2,0)} + s\bd{B}_2^{(0,1)}\mid^{(0,0)}\,,\\
%%%%%%%%%%%%
 \bd{B}_2^{(1,1)}(s,\bar{s})\ =&\ \bd{B}_2^{(1,1)}\mid^{(0,0)}\,,\\ 
 \bd{\lambda}_2^{(1,0)}(s,\bar{s})\ =&\ \bd{\lambda}_2^{(1,0)}\mid^{(0,2)} 
+ \bar{s}\bd{\lambda}_2^{(1,0)}\mid^{(0,0)}\,,\qquad
\bd{\lambda}_2^{(1,1)}(s,\bar{s})\ =\ \bd{\lambda}_2^{(1,1)}\mid^{(0,0)}\,,\\
%%%%%%
 \bar{\bd{\lambda}}_2^{(0,1)}(s,\bar{s})\ =&\ \bd{\lambda}_2^{(0,1)}\mid^{(2,0)} 
+ s \bar{\bd{\lambda}}_2^{(0,1)}\mid^{(0,0)}\,,\qquad 
%%%%%%%%%%%%%%%
\bar{\bd{\lambda}}_2^{(1,1)}(s,\bar{s})\ =\ \bar{\bd{\lambda}}_2^{(1,1)}\mid^{(0,0)}\,.
\end{align}
The explicit form of each product is found by expanding the solutions (\ref{sol 1})-(\ref{sol 4}) 
in $s$ and $\bar{s}$\,. In particular, the 2-string product satisfying (\ref{L infinity B}) is given by
\begin{align}
 \pi_1\bd{B}_2\ =&\ \pi_1\bd{B}_2^{(0,0)}\mid^{(2,2)} + \pi_1\bd{B}_2^{(1,0)}\mid^{(0,2)} 
+ \pi_1\bd{B}_2^{(0,1)}\mid^{(2,0)} + \pi_1\bd{B}_2^{(1,1)}\mid^{(0,0)}
\nonumber\\
=&\
\pi_1\bd{L}_2^B\mid^{(2,2)}
+ X\circ\pi_1\bd{L}_2^B\mid^{(0,2)}
+ \bar{X}\circ\pi_1\bd{L}_2^B\mid^{(2,0)}
%\nonumber\\
%&\
%+ \frac{1}{2}\left(X\circ\left(\bar{X}\circ\pi_1\bd{L}_2^B\mid^{(0,0)}\right)
%+\bar{X}\circ\left(X\circ\pi_1\bd{L}_2^B\mid^{(0,0)}\right)\right)\,.
+X\circ\bar{X}\circ\pi_1\bd{L}_2^B\mid^{(0,0)}\,.
\end{align}
The flow of determining each component is given in Fig.~\ref{2string comp}.

\begin{figure}[hbtp]
 \begin{center}
  \includegraphics[clip,width=15.5cm]{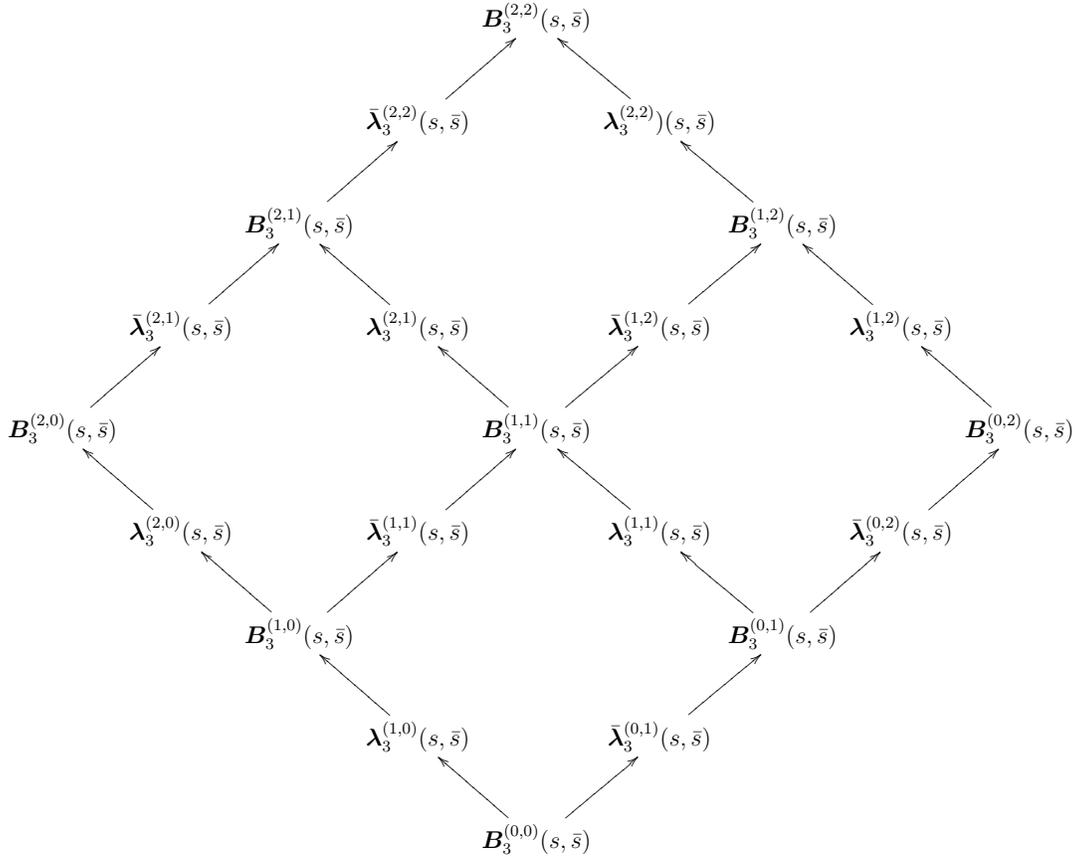}
    \caption{The flow of how 3-string (gauge) products are determined.}
      \label{3string ss}
 \end{center}
\end{figure}

Similarly, expanding the 3-string products as
\begin{align}
 \bd{B}_3\sst\ =&\ \bd{B}_3^{(0,0)}(s,\bar{s}) + 
t\left(\bd{B}_3^{(1,0)}(s,\bar{s})+\bd{B}_3^{(0,1)}(s,\bar{s})\right)
\nonumber\\
&\
+t^2\left(\bd{B}_3^{(2,0)}(s,\bar{s})+\bd{B}_3^{(1,1)}(s,\bar{s})+\bd{B}_3^{(0,2)}(s,\bar{s})\right)
\nonumber\\
&\
+t^3\left(\bd{B}_3^{(2,1)}(s,\bar{s})+\bd{B}_3^{(1,2)}(s,\bar{s})\right)
+ t^4\bd{B}_3^{(2,2)}(s,\bar{s})\,,\\
%%%%%%%%%%%%%%%%%%%
\bd{\lambda}_3\sst\ =&\ \bd{\lambda}_3^{(1,0)}(s,\bar{s}) 
+ t\left(\bd{\lambda}_3^{(2,0)}\ssb + \bd{\lambda}_3^{(1,1)}\ssb\right)
\nonumber\\
&\
+ t^2 \left(\bd{\lambda}_3^{(2,1)}\ssb + \bd{\lambda}_3^{(1,2)}\ssb\right)
+ t^3 \bd{\lambda}_3^{(2,2)}\ssb\,,\\
%%%%%%%%%%%%%%%%%%%
\bar{\bd{\lambda}}_3\sst\ =&\ \bar{\bd{\lambda}}_3^{(0,1)}(s,\bar{s}) 
+ t\left(\bar{\bd{\lambda}}_3^{(1,1)}\ssb + \bar{\bd{\lambda}}_3^{(0,2)}\ssb\right)
\nonumber\\
&\
+ t^2 \left(\bar{\bd{\lambda}}_3^{(2,1)}\ssb + \bar{\bd{\lambda}}_3^{(1,2)}\ssb\right)
+ t^3 \bar{\bd{\lambda}}_3^{(2,2)}\ssb\,,
\end{align}
the differential equations (\ref{diff eqs}) for 3-string products 
at $\mathcal{O}(t^0)$ become
\begin{subequations}\label{3string lowest}
\begin{align}
\bd{B}_3^{(1,0)}\ssb\ =&\ [\,\bd{Q}\,, \bd{\lambda}_3^{(1,0)}\ssb\,] 
+ [\,\bd{B}_2^{(0,0)}\ssb\,, \bd{\lambda}_2^{(1,0)}\ssb\,]^{11}
\nonumber\\
&\
+ s [\,\bd{B}_2^{(0,0)}\ssb\,, \bd{\lambda}_2^{(1,0)}\ssb\,]^{21}
+ \bar{s} [\,\bd{B}_2^{(0,0)}\ssb\,, \bd{\lambda}_2^{(1,0)}\ssb\,]^{12}
\nonumber\\
&\
+ s\bar{s} [\,\bd{B}_2^{(0,0)}\ssb\,, \bd{\lambda}_2^{(1,0)}\ssb\,]^{22}
+ \frac{\bar{s}}{2} [\,\bd{B}_2^{(0,0)}\ssb\,, \bd{B}_2^{(0,0)}\ssb\,]^{22}_{\Xi}\,,
\label{3string 10}\\
%%%%%%%%%%%%%%%%%%%
\bd{B}_3^{(0,1)}\ssb\ =&\ [\,\bd{Q}\,, \bar{\bd{\lambda}}_3^{(0,1)}\ssb\,] 
+ [\,\bd{B}_2^{(0,0)}\ssb\,, \bar{\bd{\lambda}}_2^{(0,1)}\ssb\,]^{11}
\nonumber\\
&\
+ s [\,\bd{B}_2^{(0,0)}\ssb\,, \bar{\bd{\lambda}}_2^{(0,1)}\ssb\,]^{21}
+ \bar{s} [\,\bd{B}_2^{(0,0)}\ssb\,, \bar{\bd{\lambda}}_2^{(0,1)}\ssb\,]^{12}
\nonumber\\
&\
+ s\bar{s} [\,\bd{B}_2^{(0,0)}\ssb\,, \bar{\bd{\lambda}}_2^{(0,1)}\ssb\,]^{22}
+ \frac{\bar{s}}{2} [\,\bd{B}_2^{(0,0)}\ssb\,, \bd{B}_2^{(0,0)}\ssb\,]^{22}_{\bar{\Xi}}\,,
\label{3string 01}\\
%%%%%%%%%%%%%%%%%%%
 \partial_s\bd{B}_3^{(0,0)}\ssb\ =&\ [\,\bd{\eta}\,, \bd{\lambda}_3^{(1,0)}\ssb\,]\,,\qquad
 \partial_{\bar{s}}\bd{B}_3^{(0,0)}\ssb\ =\ [\,\bar{\bd{\eta}}\,, \bar{\bd{\lambda}}_3\ssb\,]\,.
\label{3string ssbar}
\end{align}
\end{subequations}
Under the initial condition $\bd{B}_3^{(0,0)}(s,\bar{s})=\bd{L}_3^B(s,\bar{s})$\,,
eqs.~(\ref{3string ssbar}) are solved for $\bd{\lambda}_3^{(1,0)}\ssb$ 
and $\bar{\bd{\lambda}}_3^{(0,1)}\ssb$ as
\begin{align}
 \pi_1\bd{\lambda}_3^{(1,0)}\ssb\ =&\ \xi\circ\partial_sL_3^B\ssb\ =\
\frac{1}{4}\left(
\xi \partial_sL^B_3\ssb - \partial_s L_3^B\ssb(\xi\pi_1\wedge\id_2)\right)\,,\\
 \pi_1\bar{\bd{\lambda}}_3^{(0,1)}\ssb\ =&\ \bar{\xi}\circ\partial_{\bar{s}}L_3^B\ssb\ =\
\frac{1}{4}\left(
\bar{\xi} \partial_{\bar{s}}L^B_3\ssb - \partial_{\bar{s}} L_3^B\ssb(\bar{\xi}\pi_1\wedge\id_2)\right)\,.
\end{align}
Then, all the terms on the right-hand sides of eqs.~(\ref{3string 10}) and (\ref{3string 01}) are
given, and thus, they determine $\bd{B}_3^{(1,0)}\ssb$ and $\bd{B}_3^{(0,1)}\ssb$\,. 
We can repeat similar procedures at each order of $t$ and determine all the products
$\bd{B}_3^{(\bullet,\bullet)}(s,\bar{s})$, 
$\bd{\lambda}_3^{(\bullet,\bullet)}(s,\bar{s})$\,, 
and $\bar{\bd{\lambda}}_3^{(\bullet,\bullet)}(s,\bar{s})$\,.
In general, the equations (\ref{diff eqs}) at each order of $t$ have the form\footnote{
Depending on the picture number, $\bd{\lambda}_3^{(\bullet,\bullet)}\ssb$ 
or $\bar{\bd{\lambda}}_3^{(\bullet,\bullet)}\ssb$ is missing in eq.~(\ref{eq B}).}
\begin{subequations} 
\begin{align}
% \bd{B}_3^{(\bullet,\bullet)}\ssb\ =&\ [\,\bd{Q}\,, \llb_3^{(\bullet,\bullet)}\ssb\,] + (\textrm{kown\ terms})\,,
 \bd{B}_3^{(\bullet,\bullet)}\ssb\ =&\ [\,\bd{Q}\,, \bd{\lambda}_3^{(\bullet,\bullet)}\ssb
+ \bar{\bd{\lambda}}_3^{(\bullet,\bullet)}\ssb\,] + \cdots\,,
%(\textrm{kown\ terms})\,,
\label{eq B}\\
 [\,\bd{\eta}\,, \bd{\lambda}^{(\bullet,\bullet)}_3\ssb\,]\ =&\ \cdots\,,\qquad
%(\textrm{known\ terms})\,,\qquad
 [\,\bar{\bd{\eta}}\,, \bar{\bd{\lambda}}^{(\bullet,\bullet)}_3\ssb\,]\ =\ \cdots\,,
%(\textrm{known\ terms})\,.
\label{gauge ast}
\end{align}
\end{subequations}
where $\cdots$ denotes the terms including only the products already given in the lower order.
We can solve eqs.~(\ref{gauge ast}) for $\bd{\lambda}_3^{(\bullet,\bullet)}\ssb$ 
or $\bar{\bd{\lambda}}_3^{(\bullet,\bullet)}\ssb$\,, and then
eq.~(\ref{eq B}) determine $\bd{B}_3^{(\bullet,\bullet)}\ssb$\,.
We give the flow of how the 3-string (gauge) products are determined
in Fig.~\ref{3string ss}. Their expansion in $s$ and $\bar{s}$ and the flow of how each component,
(gauge) products with specific cyclic Ramond number, 
are determined are given in Appendix~\ref{explicit B3}.
In principle, we can repeat a similar procedure for higher-string products 
$\bd{B}_n^{(\bullet,\bullet)}(s,\bar{s})$\,,
$\bd{\lambda}_n^{(\bullet,\bullet)}(s,\bar{s})$\,, and
$\bar{\bd{\lambda}}_n^{(\bullet,\bullet)}(s,\bar{s})$
in ascending order of $n$ and determine all the string and gauge products $\bd{B}\sst$\,, $\bd{\lambda}\sst$ 
and $\bar{\bd{\lambda}}\sst$\,.

Before closing this section, we rewrite
the (extended) $L_\infty$-relations (\ref{ext L infinity}) 
and the differential equations (\ref{diff eqs}) as
\begin{subequations}\label{alt ext L infinity}  
\begin{align}
&\ [\bd{Q}\,, \bd{B}_{n+2}(s,\bar{s},t)] 
+ \sum_{m=0}^{n-1}\bd{B}_{m+2}(s,\bar{s},t)\Big(\pi(s,\bar{s},t)\pi_1\bd{B}_{n-m+1}(s,\bar{s},t)
\wedge\id_{m+1}\Big)\ =\ 0\,,\\
%%%%%%%%%%%%%%%%%%%
&\ [\bd{\eta}\,, \bd{B}_{n+2}(s,\bar{s},t)] 
+ \sum_{m=0}^{n-1}\bd{B}_{m+2}(s,\bar{s},t)\Big(\pi(t)\pi_1\bd{B}_{n-m+1}(s,\bar{s},t)
\wedge\id_{m+1}\Big)\ =\ 0\,,\\
%%%%%%%%%%%%%%%%%%%
&\ [\bar{\bd{\eta}}\,, \bd{B}_{n+2}(s,\bar{s},t)] 
+ \sum_{m=0}^{n-1}\bd{B}_{m+2}(s,\bar{s},t)\Big(\bar{\pi}(t)\pi_1\bd{B}_{n-m+1}(s,\bar{s},t)
\wedge\id_{m+1}\Big)\ =\ 0\,,
\end{align}
\end{subequations}
and
\begin{subequations}\label{alt diff eq}  
\begin{align}
 \partial_t\bd{B}_{n+2}(s,\bar{s},t)\ =&\ [\bd{Q}\,, (\bd{\lambda}+\bar{\bd{\lambda}})_{n+2}(s,\bar{s},t)]
\nonumber\\
&\
+ \sum_{m=0}^{n-1}\Big(\bd{B}_{m+2}(s,\bar{s},t)\left(\pi(s,\bar{s},t)\pi_1(\bd{\lambda}+\bar{\bd{\lambda}})_{n-m+1}
(s,\bar{s},t)\wedge\id_{m+1}\right)
\nonumber\\
&\hspace{15mm}
-(\bd{\lambda}+\bar{\bd{\lambda}})_{m+2}(s,\bar{s},t)\left(\pi(s,\bar{s},t)\pi_1\bd{B}_{n-m+1}(s,\bar{s},t)
\wedge\id_{m+1}\right)\Big)
\nonumber\\
&\
+ \sum_{m=0}^{n-1}\bd{B}_{m+2}(s,\bar{s},t)
 \left((s\bar{\Xi}+\bar{s}\Xi)\pi_1^{(1,1)}\bd{B}_{n-m+1}(s,\bar{s},t)\wedge\id_{m+1}\right)
\,,\\
%%%%%%%%%%%%%%%
\partial_s\bd{B}_{n+2}(s,\bar{s},t)\ =&\
[\bd{\eta}\,, \bd{\lambda}_{n+2}(s,\bar{s},t)]
\nonumber\\
&
+ \sum_{m=0}^{n-1}\Big(\bd{B}_{m+2}(s,\bar{s},t)\left(\pi(t)\pi_1(\bd{\lambda}+\bar{\bd{\lambda}})_{n-m+1}
(s,\bar{s},t)\wedge\id_{m+1}\right)
\nonumber\\
&\hspace{15mm}
-(\bd{\lambda}+\bar{\bd{\lambda}})_{m+2}(s,\bar{s},t)\left(\pi(t)\pi_1\bd{B}_{n-m+1}(s,\bar{s},t)
\wedge\id_{m+1}\right)\Big)
\,,\\
%%%%%%%%%%%%%%%
\partial_{\bar{s}}\bd{B}_{n+2}(s,\bar{s},t)\ =&\
[\bar{\bd{\eta}}\,, \bar{\bd{\lambda}}_{n+2}(s,\bar{s},t)]
\nonumber\\
&
+ \sum_{m=0}^{n-1}\Big(\bd{B}_{m+2}(s,\bar{s},t)\left(\bar{\pi}(t)\pi_1(\bd{\lambda}+\bar{\bd{\lambda}})_{n-m+1}
(s,\bar{s},t)\wedge\id_{m+1}\right)
\nonumber\\
&\hspace{15mm}
-(\bd{\lambda}+\bar{\bd{\lambda}})_{m+2}(s,\bar{s},t)\left(\bar{\pi}(t)\pi_1\bd{B}_{n-m+1}(s,\bar{s},t)
\wedge\id_{m+1}\right)\Big)\,,
\end{align}
\end{subequations}
where
\begin{equation}
\begin{split}
&\ \pi(s,\bar{s},t)\ =\ \pi^{(0,0)}+s\pi^{(1,0)}+\bar{s}\pi^{(0,1)}
+\big(t\left(s\bar{X}+\bar{s}X\right)+s\bar{s}\big)\pi^{(1,1)}\,,\\
&\ \pi(t)\ =\ -t\left(\pi^{(1,0)}+\frac{t}{2}\bar{X}\pi^{(1,1)}\right)\,,\qquad
\bar{\pi}(t)\ =\ -t\left(\pi^{(0,1)}+\frac{t}{2}X\pi^{(1,1)}\right)\,.
\end{split}
\end{equation}
These alternative forms are convenient for use in the next section \cite{Kunitomo:2020xrl}.

\sectiono{Tree-level S-matrix}\label{sec S-matrix}

Using HPT, we can show that the tree-level S-matrix derived from (super)string field theory
agrees with that calculated using the first-quantization method 
\cite{Kajiura:2003ax,Konopka:2015tta,Kunitomo:2020xrl}.
We show in this section that the new prescription proposed in the previous section simplifies 
this proof for the type II superstring field theory.

As a preparation, let us define the on-shell subspace,
\begin{equation}
 \mathcal{H}^p\ =\ \{\Phi\in\mathcal{H}^{res} \mid L_0^+\Phi_{\NSNS} =  G_0\Phi_{\RNS} =
\bar{G}_0\Phi_{\NSR} = G_0\Phi_{\RR} = \bar{G}_0\Phi_{\RR} = 0\}\,,
\end{equation}
and the BRST invariant projection operator,
\begin{equation}
 P_0\ :\ \mathcal{H}^{res}\longrightarrow\mathcal{H}^p\,,\qquad
P_0^2=P_0\,,\qquad [Q,P_0]=0\,.
\end{equation}
The homotopy operator $Q^+$ of $Q$ satisfying
\begin{align}
 Q^+Q+QQ^++P_0\ =\ \id\,,\\
Q^+P_0 = P_0Q^+ = Q^+Q^+ = 0\,,
\end{align}
is defined by
\begin{equation}
 Q^+\ =\ \frac{1}{L_0^+}b_0^+(1-P_0)\,,
\end{equation}
and provides a Hodge-Kodaira decomposition of $\mathcal{H}^{res}$\,,
\begin{equation}
 \mathcal{H}^{res}\ =\ \mathcal{H}^p + \mathcal{H}^t + \mathcal{H}^u\,,\\
\end{equation}
with
\begin{equation}
\mathcal{H}^p  = P_0\mathcal{H}^{res}\,,\qquad
\mathcal{H}^t  = QQ^+\mathcal{H}^{res}\,,\qquad
\mathcal{H}^u = Q^+Q\mathcal{H}^{res}\,.
\end{equation}
%This decomposition 
It is compatible with $\Omega$\,:
\begin{equation}
 \Omega(\mathcal{H}^p\,,\mathcal{H}^u)\ =\ \Omega(\mathcal{H}^u\,,\mathcal{H}^u)\ =\ 0\,.
\end{equation}
 
Consider two chain complexes, $(\mathcal{H}^{res},Q)$ and $(\mathcal{H}^p,QP_0)$\,,
with chain maps
\begin{equation}
 p\ =\ P_0\,:\, \mathcal{H}^{res}\ \twoheadrightarrow\ \mathcal{H}^p\,,\qquad
 i\ =\ P_0\,:\,\mathcal{H}^p\ \hookrightarrow\ \mathcal{H}^{res}\,,
\label{chain map}
\end{equation}
satisfying the relations
\begin{equation}
 pi\ =\ P_0\,,\qquad ip\ =\ \id -Q^+Q-QQ^+\,.
\end{equation}
Under the SR gauge condition $\psi=0$\,, $\mathcal{H}^t=\emptyset$\,,
and the Hilbert space of the quantum string field $\phi$ is
decomposed to the on-shell and off-shell components\,:
\begin{equation}
 \phi\ =\ P_0\phi + (1-P_0)\phi\ \in\ \mathcal{H}_0+\overline{\mathcal{H}}_0\,,
\end{equation}
where\footnote{
To be precise, we must remove the ghost number constraint for the quantum string field, 
but components with non-zero space-time ghost number do not contribute the tree-level S-matrix.
}
\begin{equation}
 \mathcal{H}_0 = \mathcal{H}^p\cap\mathcal{H}_{SR} = P_0\mathcal{H}_{SR}\,,\qquad
\overline{\mathcal{H}}_0 = \mathcal{H}^u \cap\mathcal{H}_{SR} = (1-P_0)\mathcal{H}_{SR}\,.
\end{equation}
The chain complex $(\mathcal{H}^p, QP_0)$\,, with the SR gauge condition, defines
the relative BRST cohomology.
Lift these (equivalence) data to the chain complexes 
$(\mathcal{SH}^{res},\bd{Q})$ and $(\mathcal{SH}^p,\bd{QP})$ with the chain maps
\begin{equation}
 \hat{\bd{p}}\ =\ \hat{\bd{P}}\,:\, \mathcal{SH}^{res}\ \twoheadrightarrow\ \mathcal{SH}^p\,,\qquad
 \hat{\bd{i}}\ =\ \hat{\bd{P}}\,:\,\mathcal{SH}^p\ \hookrightarrow\ \mathcal{SH}^{res}\,,
\label{chain map SH}
\end{equation}
satisfying
\begin{equation}
 \hat{\bd{p}}\hat{\bd{i}}\ =\ \hat{\bd{P}}\,,\qquad
 \hat{\bd{i}}\hat{\bd{p}}\ =\ \hat{\bd{I}}+\bd{H}\bd{Q}+\bd{Q}\bd{H}\,,
\end{equation}
where
%\begin{equation}
%\xymatrix{{} \ar@(ul,dl)[]_{-h}}
%(\mathcal{H}^{res}, d=Q)
%\xymatrix{
%{} \ar@<1 ex>[r]^{p} & {} \ar@<1 ex>[l]^{i}}
%(\mathcal{H}^p, D=QP_0)
%\end{equation}
\begin{align}
\hat{\bd{P}}\ =&\ \sum_{n=0}^\infty P_n\ =\ \sum_{n=0}^\infty\frac{1}{n!}\underbrace{P_0\wedge\cdots\wedge P_0}_n\,,\\
\hat{\bd{I}}\ =&\ \sum_{n=0}^\infty\ =\ \sum_{n=0}^\infty \frac{1}{n!}\underbrace{\id_1\wedge\cdots\id_1}_n\,,\\
 \bd{H}\ =&\ \sum_{r,s=0}^\infty\frac{1}{(r+s+1)!}(-Q^+)\wedge\underbrace{\id_1\wedge\cdots\wedge\id_1}_r\wedge
\underbrace{P_0\wedge\cdots\wedge P_0}_s\,.
\end{align}
The projection, identity, and homotopy operators, $\hat{\bd{P}}$\,, $\hat{\bd{I}}$\,,
and $\bd{H}$\,, satisfy the relations
 \begin{align}
&\ \hat{\bd{P}}\ =\ \hat{\bd{I}} + \bd{H}\bd{Q} + \bd{Q}\bd{H}\,,\\
&\ \hat{\bd{P}}^2\ =\ \hat{\bd{P}}\,,\qquad
[\,\bd{Q}\,, \hat{\bd{P}}\,]\ =\ 0\,,\\
&\ \bd{H}\hat{\bd{P}}\ =\ \hat{\bd{P}}\bd{H}\ =\ \bd{H}\bd{H}\ =\ 0\,.
 \end{align}
If we perturb $\bd{Q}$ by $\bd{L}_{int}=\sum_{n=0}^\infty\bd{L}_{n+2}$ so that
$(\bd{Q}+\bd{L}_{int})^2=0$\,, the homological perturbation lemma tells us
that chain complexes and chain maps are deformed as
\begin{align}
(\mathcal{SH}^{res}, \bd{L}=\bd{Q}+\bd{L}_{int})
\xymatrix{
{} \ar@<1 ex>[r]^{\bd{p}'} & {} \ar@<1 ex>[l]^{\bd{i}'}}
(\mathcal{SH}^p, \bd{S}=\bd{QP}+\bd{S}_{int})\,, 
\end{align}
with
\begin{align}
 \hat{\bd{p}}'\ =&\ \hat{\bd{P}}(\hat{\bd{I}}-\bd{L}_{int}\bd{H})^{-1}\,,\\
\hat{\bd{i}}'\ =&\ (\hat{\bd{I}}-\bd{H}\bd{L}_{int})^{-1}\hat{\bd{P}}\,,
\label{def i'}\\
\bd{S}_{int}\ =&\ \hat{\bd{P}}\bd{L}_{int}(\hat{\bd{I}}-\bd{H}\bd{L}_{int})^{-1}\hat{\bd{P}}\,.
\end{align}
Here, we defined $(\hat{\bd{I}}-\bd{\mathcal{O}})^{-1}$ by the formal series:
\begin{equation}
 (\hat{\bd{I}}-\bd{\mathcal{O}})^{-1}\ =\ \sum_{n=0}^\infty\mathcal{O}^n\,.
\end{equation}
For convenience, introduce $\bd{\Sigma}$ with
\begin{equation}
\pi_1\bd{S}_{int}\ =\ 
\mathcal{G}P_0\pi_1\bd{\Sigma}\,,\qquad
 \bd{\Sigma}\ 
=\ \bd{B}\hat{\bd{F}}\hat{\bd{i}}'\,.
\end{equation}
The multi-linear representation,
\begin{equation}
 \langle S|\ =\ \langle\Omega|P_0\otimes \pi_1\bd{S}_{int}\
=\ \langle\omega_l|\xi\bar{\xi}P_0\otimes P_0\pi_1\bd{\Sigma}\,,
\label{S matrix}
\end{equation}
defines a map from $\mathcal{H}^p\otimes\mathcal{SH}^p$ to $\mathbb{C}$\,.
The \textit{total} S-matrix (at the tree level) is given by $\langle S|$ by restricting 
the external states onto the gauge-fixed states, that is,
$\mathcal{H}^p\otimes\mathcal{SH}^p$ to $\mathcal{H}_0\otimes\mathcal{SH}_0$\,.
If we expand $\bd{\Sigma}$ in the number of inputs, 
$\bd{\Sigma}=\sum_{n=0}^\infty \bd{\Sigma}_{n+2}$\,, it induces the expansion of 
the S-matrix in the number of external states:
%We expand $\bd{\Sigma}$ in the number of inputs as $\bd{\Sigma}=\sum_{n=0}^\infty \bd{\Sigma}_{n+2}$\,,
%which gives the expansion of the S-matrix $\langle S|$ in the number of external states:
\begin{equation}
 \langle S|\ =\ \sum_{n=0}^\infty \langle S_{n+3}|\ \,,\qquad \langle S_{n+3}|\ 
=\ %\sum_{n=0}^\infty
\langle\omega_l|\xi\bar{\xi}P_0\otimes P_0\pi_1\bd{\Sigma}_{n+2}\,.
\label{S-matrix}
\end{equation}
Each term with specific number of external states 
is further classified by the number of external Ramond states:
\begin{align}
\langle S_{n+3}|\ =&\
\sum_{0\le r,\bar{r}\le (n+3)/2} \langle S_{n+3}|^{(2r,2\bar{r})}
\nonumber\\
=&\ \sum_{0\le r,\bar{r}\le (n+3)/2}
\langle\omega_l|\xi\bar{\xi}P_0\otimes P_0\pi_1\bd{\Sigma}_{n+2}|^{(2r,2\bar{r})}\,.
\end{align}
For example, $\langle S_3|$ contains four terms:
\begin{equation}
 \langle S_3|\ =\
\langle S_3|^{(0,0)}
+ \langle S_3|^{(0,2)}
+ \langle S_3|^{(2,0)}
+ \langle S_3| ^{(2,2)}\,.
\end{equation}

The first, second, and third terms
%, $\langle S_3|^{(0,0)}$\,, $\langle S_3|^{(0,2)}$\,, and $\langle S_3|^{(2,0)}$ 
give $(\textrm{NS-NS})^3$\,,
$(\textrm{NS-NS})\textrm{-}(\textrm{R-NS})^2$\,, and $(\textrm{NS-NS})\textrm{-}(\textrm{NS-R})^2$
amplitudes, respectively. 
The fourth term, on the other hand, 
includes two amplitudes $(\textrm{NS-NS})$-$(\textrm{R-R})^2$ and 
$(\textrm{R-NS})\textrm{-}(\textrm{NS-R})\textrm{-}(\textrm{R-R})$\,,
which cannot be distinguished by the  Ramond number of the external states\footnote{
To be precise, the second, third, or fourth term contains two different representations for 
an amplitude due to the asymmetry that distinguishes one external state (output) from the others (inputs).}.

Since the Hilbert space $\mathcal{H}_0$ still contains unphysical states in general, 
the \textit{physical} S-matrix is defined by projecting it onto the physical subspace, that is,
by taking physical states 
defined by the relative BRST cohomology, $\mathcal{H}_Q=\textrm{Ker}\, Q /\textrm{Im}\,Q
\subset\mathcal{H}_0$\,, as external states:
\begin{equation}
 \langle S^{phys}|\ =\ \langle S|\hat{\bd{\mathcal{P}}}\,,
\end{equation}
with
\begin{equation}
\hat{\bd{\mathcal{P}}}\ =\ \sum_{n=0}^\infty\frac{1}{n!}
\underbrace{\mathcal{P}\wedge\cdots\wedge\mathcal{P}}_n\,,\qquad
 \mathcal{P}\ :\ \mathcal{H}_0\ \longrightarrow\ \mathcal{H}_Q\,.
\end{equation}
The unitarity of the physical S-matrix is guaranteed by the BRST invariance,
$[\bd{Q},\bd{S}_{int}]=0$\,.

By using the relations of the cohomomorphisms
\begin{equation}
 \pi_1\hat{\bd{F}}\ =\ \pi_1\id + \mathfrak{S}\pi_1\bd{B}\hat{\bd{F}}\,,\qquad
\pi_1\hat{\bd{i}}'\ =\ P_0\pi_1-Q^+\mathcal{G}\pi_1\bd{B}\hat{\bd{F}}\hat{\bd{i}}'
%\bd{\Sigma}\,,
\end{equation}
following from the definitions (\ref{def F}) of $\hat{\bd{F}}^{-1}$
and  (\ref{def i'}) of $\hat{\bd{i}}'$, respectively,
we can show that the Dyson-Schwinger (DS) equation
for $\bd{\Sigma}$\,,
\begin{equation}
 \pi_1\bd{\Sigma}\ 
=\ \sum_{n=0}^\infty\pi_1\bd{B}_{n+2}
\left(\frac{1}{(n+2)!}\big(P_0\pi_1-
%\Delta
(Q^+\mathcal{G} - \mathfrak{S})
\pi_1\bd{\Sigma}\big)^{\wedge(n+2)}\right)\,,
\label{DS eq}
\end{equation}
holds.
By expanding $\bd{\Sigma}$ in the number of inputs, %$\bd{\Sigma}=\sum_{n=0}^\infty\bd{\Sigma}_{n+2}$\,,
it becomes the recurrence relation 
\begin{equation}
  \pi_1\bd{\Sigma}_{n+2}\ 
=\ \sum_{m=0}^n\pi_1\bd{B}_{m+2}
\left(\frac{1}{(m+2)!}\big(P_0\pi_1-
%\Delta
(Q^+\mathcal{G} - \mathfrak{S})\pi_1
\sum_{l=0}^{n-m-1}\bd{\Sigma}_{l+2}\big)^{\wedge(m+2)}\right)\pi_{n+2}\,,
\label{DS eq rec}
\end{equation}
which determine $\bd{\Sigma}_{n+2}$ recursively. % and thus define $\bd{\Sigma}$\,.
We extend it with three parameters $s$\,, $\bar{s}$\,, and $t$ to 
the extended DS equation,
\begin{align}
 &\pi_1\bd{\Sigma}\sst_{n+2}\ 
\nonumber\\
&\hspace{5mm} =\ \sum_{m=0}^n\pi_1\bd{B}_{m+2}\sst
\left(\frac{1}{(m+2)!}\big(P_0\pi_1-
\Delta\sst
%(Q^+\mathcal{G} - \mathfrak{S})
\pi_1\sum_{l=0}^{n-m-1}\bd{\Sigma}_{l+2}\sst\big)^{\wedge(m+2)}\right)\,,
\label{ext DS eq}
\end{align}
which reduces to the DS equation (\ref{DS eq rec}) at $\sst=(0,0,1)$\,.
Here, $\Delta\sst$ defined by 
\begin{equation}
\Delta(s,\bar{s},t)=Q^+\mathcal{G}(s,\bar{s},t) - \mathfrak{S}(t)\,,
\label{def Delta} 
\end{equation}
with
\begin{align}
%\Delta(s,\bar{s},t)\, =&\, Q^+\mathcal{G}(s,\bar{s},t) - \mathfrak{S}(t)\,,
%\label{def Delta} \\
\mathcal{G}(s,\bar{s},t)\,=&\,
\pi^{(0,0)}+(tX+s)\pi^{(1,0)} + (t\bar{X}+\bar{s})\pi^{(0,1)}
+ \left(t^2X\bar{X}+t(s\bar{X}+\bar{s}X)+s\bar{s}\right)\pi^{(1,1)}\,,\\
 \mathfrak{S}(t)\, =&\,
t\, \Xi\pi^{(1,0)}+t\,\bar{\Xi}\pi^{(0,1)} 
+ \frac{t^2}{2}(\Xi\bar{X}+X\bar{\Xi})\pi^{(1,1)}\,,
\end{align}
%The $\Delta\sst$ defined by (\ref{def Delta}) 
was determined to satisfy the relations
%it satisfies the relations
\begin{subequations}\label{Delta relations} 
 \begin{align}
&\ [Q\,, \Delta(s,\bar{s},t)]\ =\ \pi(s,\bar{s},t)-P_0\mathcal{G}(s,\bar{s},t)\,,\\
&\ [\eta\,, \Delta(s,\bar{s},t)]\ =\ \pi(t)\,,\qquad
 [\bar{\eta}\,, \Delta(s,\bar{s},t)]\ =\ \bar{\pi}(t)\,,
\end{align}
and
\begin{align}
 \partial_t\Delta(s,\bar{s},t)\ =&\ 
- \left[Q,Q^+\left(\partial_t\mathfrak{S}(t)+(s\bar{\Xi}+\bar{s}\Xi)\pi^{(1,1)}\right)\right]
+ (s\bar{\Xi}+\bar{s}\Xi)\pi^{(1,1)} 
\nonumber\\
&\
+ P_0\left(\partial_t\mathfrak{S}(t)+(s\bar{\Xi}+\bar{s}\Xi)\pi^{(1,1)}\right)\,,\\
%%%%%%%%%%%%%%%
 \partial_s\Delta(s,\bar{s},t)\ =&\ 
- \left[\eta\,, Q^+\left(\partial_t\mathfrak{S}(t)+(s\bar{\Xi}+\bar{s}\Xi)\pi^{(1,1)}\right)\right]\,,\\
%%%%%%%%%%%%%%%
 \partial_{\bar{s}}\Delta(s,\bar{s},t)\ =&\ 
- \left[\bar{\eta}\,, Q^+\left(\partial_t\mathfrak{S}(t)+(s\bar{\Xi}+\bar{s}\Xi)\pi^{(1,1)}\right)\right]\,.
\end{align}
\end{subequations}
These relations (\ref{Delta relations}) make the $\bd{\Sigma}\sst$ satisfying
(\ref{ext DS eq}) is BRST invariant and is in the small Hilbert space:
\begin{subequations}  
\begin{equation}
 [\,\bd{Q}\,, \bd{\Sigma}\sst\,]\ =\ 0\,,
\end{equation}
\begin{equation}
[\,\bd{\eta}\,, \bd{\Sigma}\sst\,]\ =\ 0\,,\qquad
[\,\bar{\bd{\eta}}\,, \bd{\Sigma}\sst\,]\ =\ 0\,,
\end{equation}
\end{subequations}
which can be shown, similar to the case in \cite{Kunitomo:2020xrl},
using the alternative form of the (extended) $L_\infty$-relations 
(\ref{alt ext L infinity}) and the fact that generic intermediate 
states are off-shell.
Next, differentiating eq.~(\ref{ext DS eq}) by parameters, 
we obtain the relations
\begin{subequations}\label{diff rel} 
 \begin{equation}
 \pi_1\partial_t\bd{\Sigma}\sst\ =\ \pi_1[\,\bd{Q}\,, \bd{\rho}\sst\,]\,, 
\end{equation}
\begin{equation}
 \pi_1\partial_s\bd{\Sigma}\sst\ =\ \pi_1[\,\bd{\eta}\,, \bd{\rho}\sst\,]\,,\qquad
 \pi_1\partial_{\bar{s}}\bd{\Sigma}\sst\ =\ \pi_1[\,\bar{\bd{\eta}}\,, \bd{\rho}\sst\,]\,,
\end{equation}
\end{subequations}
%\begin{align}
%\begin{split}
% \pi_1\partial_t\bd{\Sigma}\sst\ =&\ \pi_1[\,\bd{Q}\,, \bd{\rho}\,]\,,\\
% \pi_1\partial_s\bd{\Sigma}\sst\ =&\ \pi_1[\,\bd{\eta}\,, \bd{\rho}\,]\,,\qquad
% \pi_1\partial_{\bar{s}}\bd{\Sigma}\sst\ =\ \pi_1[\,\bar{\bd{\eta}}\,, \bd{\rho}\,]\,,
% \end{split}
%\end{align}
using the (alternative form of)
the differential equations (\ref{alt diff eq}),
where $\bd{\rho}\sst$ is the degree even map determined by the recurrence relation
\begin{align}
 \pi_1\bd{\rho}_{n+2}\sst\ =&\ \sum_{m=0}^n\llb_{m+2}\sst\Big(D_{m+2}\sst\Big) P_{n+2}\pi_{n+2}
\nonumber\\
&\
- \sum_{m=0}^n\pi_1\bd{B}_{m+2}\sst\Big(D_{m+1}\sst\wedge\pi_1\bd{E}\sst\Big) P_{n+2}\pi_{n+2}\,,
\end{align}
with
\begin{equation}
 D_{M}\sst\ =\ \frac{1}{M!}\big(P_0\pi_1-
\Delta\sst\pi_1\bd{\Sigma}\sst\big)^{\wedge M}\,,
\end{equation}
\begin{equation}
 \pi_1\bd{E}\sst\ =\ \Delta\sst\pi_1\bd{\rho}+Q^+(\partial_t\mathfrak{S}
+(s\bar{\Xi}+\bar{s}\Xi)\pi^{(1,1)})\pi_1\bd{\Sigma}\sst\,.
\end{equation}
Combining the relations (\ref{diff rel}), we can derive the key relation
\begin{equation}
 \pi_1\partial_t\bd{\Sigma}\sst - X_0\circ\pi_1\partial_s\bd{\Sigma}\sst 
- \bar{X}_0\circ\pi_1\partial_{\bar{s}}\bd{\Sigma}\sst\ =\ 
[\,\bd{Q}\,, [\,\bd{\eta}\,,[\,\bar{\bd{\eta}}\,,\bd{T}\sst\,]\,]\,]\,,
\label{key eq}
\end{equation}
with $\pi_1\bd{T}\sst=\bar{\xi}\circ\xi\circ\pi_1\bd{\rho}\sst$\,.

We can also extend the S-matrix using 
$\bd{\Sigma}\sst=\sum_{n=0}^\infty\bd{\Sigma}_{n+2}\sst$ as
\begin{equation}
 \langle S\sst|\ =\ \sum_{n=0}^\infty \langle S_{n+3}\sst|\ =\ 
\sum_{n=0}^\infty \langle\omega_l|\xi\bar{\xi}P_0\otimes P_0\pi_1\bd{\Sigma}_{n+2}\sst\,,
\end{equation}
which reduces the S-matrix (\ref{S-matrix}) at $\sst=(0,0,1)$\,. 
The extended S-matrix element %$(n+3)$-string amplitudes 
$\langle S_{n+3}\sst|$ is expanded in the parameters as\footnote{
We consider that $\langle S_{n+3}^{(p,\bar{p})}|^{(2r,2\bar{r})}$ with the parameters
outside the range $(0,0)\le (p, \bar{p})\le (n+1-r,n+1-\bar{r})$\,, $ 0\le 2r, 2\bar{r}\le n+3$
is equal to zero.}
\begin{subequations}  
\begin{align}
 \langle S_{n+3}\sst|\ =&\ \sum_{m=0}^{}s^m\bar{s}^{\bar{m}}\langle S_{n+3}^{[m,\bar{m}]}(t)|\,,\\
\langle S_{n+3}^{[m,\bar{m}]}(t)|\ =&\ 
\sum_{p=0}^{n-m+1}\sum_{\bar{p}=0}^{n-\bar{m}+1}t^{p+\bar{p}}
\langle S_{n+3}^{(p,\bar{p})}|^{(2(n-m-p+1),2(n-\bar{m}-\bar{p}+1))}\,.
\end{align}
\end{subequations}
The key relation (\ref{key eq}) induces the equation
for the extended S-matrix, 
\begin{align}
%&\ 
\partial_t\langle S\sst| 
- \partial_s\langle S\sst|(X_0)_{cyc} 
&- \partial_{\bar{s}}\langle S\sst|(\bar{X}_0)_{cyc}\
\nonumber\\
&=\  \langle\omega_l|\xi\bar{\xi}P_0\otimes P_0\pi_1[\bd{Q},[\bd{\eta},\bar{\bd{\eta}},\bd{T}\sst]]\,,
\label{key eq S}
\end{align}
where
\begin{equation}
 \langle S\sst|(\mathcal{X}_0)_{cyc}\ =\ 
\sum_{n=0}^\infty\langle S_{n+3}\sst|(\mathcal{X}_0\otimes\id_{n+2}
+\id\otimes\id_{n+1}\wedge \mathcal{X}_0\pi_1)\,,
\end{equation}
for $\mathcal{X}_0=X_0$ or $\bar{X}_0$\,.
Note that the right hand side of (\ref{key eq S}) does not contribute 
to the physical S-matrix. 
For each amplitude, we have
\begin{align}
%%%%%%%%%%%%%%
(p+\bar{p})\langle S_{n+3}^{(p,\bar{p})}|^{(2r,2\bar{r})}\ %\hat{\bd{\mathcal{P}}}\ 
=&\ (n-p+1-r)\langle S_{n+3}^{(p-1,\bar{p})}|^{(2r,2\bar{r})}
(X_0)_{cyc}%\hat{\bd{\mathcal{P}}}
\nonumber\\
&\hspace{1.5cm}
+ (n-\bar{p}+1-\bar{r})\langle S_{n+3}^{(p,\bar{p}-1)}|^{(2r,2\bar{r})}
(\bar{X}_0)_{cyc} + \cdots\,,
%\hat{\bd{\mathcal{P}}}\,.
\label{key eq S comp}
%\nonumber\\
%&\hspace{-30mm}
%(0,0)\le (p, \bar{p})\le (n+1-r,n+1-\bar{r}), 0\le 2r, 2\bar{r}\le n+3
\end{align}
%By using this relation, the amplitude 
%$ \langle S_{n+3}^{(p,\bar{p})}|^{(2r,2\bar{r})}$ can be written using
%the \textit{bosonic} amplitude
%$ \langle S_{n+3}^{(0,0)}|^{(2r,2\bar{r})}$ as follows.
%% as\footnote{
%%A derivation of this relation is give in Applendix~\ref{comb}.}
where dots on the right-hand side represent the terms vanishing on
the physical subspace $\mathcal{H}_Q$\,.
Let us apply this relation %(\ref{key eq S comp}) 
on $(p+\bar{p})!\langle S_{n+3}^{(p,\bar{p})}|^{(2r,2\bar{r})}$\,, repeatedly.
By using the relation once, we can reduce $p$ or $\bar{p}$ by 1
with the factor $(n-r-p+2)$ or $(n-\bar{r}-\bar{p}+2)$\,, respectively.
By repeatedly using the relation $(p+\bar{p})$ times, 
we eventually reach $\langle S_{n+3}^{(0,0)}|^{(2r,2\bar{r})}$
through ${}_{p+\bar{p}}C_p$ paths. 
The number of paths can be obtained by counting the places to reduce $p$
in the $(p+\bar{p})$ steps. The coefficients coming from the relation
are the same for all the paths, and equal to 
\begin{equation}
 \frac{(n-r+1)!}{(n-r-p+1)!} \frac{(n-\bar{r}+1)!}{(n-\bar{r}-\bar{p}+1)!}\,.
\end{equation}
Thus, we eventually find that
%\begin{equation}
% (p+\bar{p})!\langle S_{n+3}^{(p,\bar{p})}|^{(2r,2\bar{r})}\ =\
%\frac{(p+\bar{p})!}{p!\bar{p}!}
%\frac{(n-r+1)!}{(n-r-p+1)!} \frac{(n-\bar{r}+1)!}{(n-\bar{r}-\bar{p}+1)!}
%\langle S_{n+3}^{(0,0)}|^{(2r,2\bar{r})}\,.
%\end{equation}
\begin{equation}
 \langle S_{n+3}^{(p,\bar{p})}|^{(2r,2\bar{r})}\ =\
\begin{pmatrix}
 n-r+1 \\  p 
\end{pmatrix} 
\begin{pmatrix}
 n-\bar{r}+1 \\ \bar{p} 
\end{pmatrix} 
\langle S_{n+3}^{(0,0)}|^{(2r,2\bar{r})}(X_0)_{cyc}^p(\bar{X}_0)_{cyc}^{\bar{p}}\,,
\end{equation}
except for the terms vanishing on $\mathcal{H}_Q$\,.
Then, the physical amplitudes obtained by projecting 
$\langle S_{n+3}^{(p,\bar{p})}|^{(2r,2\bar{r})}$ with 
$(p,\bar{p})=(n-r+1,n-\bar{r}+1)$ onto $\mathcal{H}_Q$
is written as
\begin{equation}
 \langle S_{n+3}^{phys}|^{(2r,2\bar{r})}\
=\ \langle (S_B)_{n+3}^{phys}|^{(2r,2\bar{r})}(X_0)_{cyc}^{n-r+1}(\bar{X}_0)_{cyc}^{n-\bar{r}+1}\,,
\end{equation}
where $\langle (S_B)^{phys}_{n+3}|^{(2r,2\bar{r})}$ in the right-hand side is
the \textit{bosonic} physical amplitude defined by
\begin{equation}
 \langle (S_B)^{phys}_{n+3}|^{(2r,2\bar{r})}\ 
=\ \langle S_{n+3}^{(0,0)}|^{(2r,2\bar{r})}
\hat{\bd{\mathcal{P}}}\,.
\end{equation}
Since the physical amplitude is independent of how we insert PCO's, 
we can appropriately deform and move them to agree with that obtained 
using the first-quantization method.

\sectiono{Gauge-invariant action in WZW-like formulation}\label{sec WZW}

Using an alternative method, symmetric construction,
we have constructed the string products (interactions) with $L_\infty$ structure
for the type II superstring.
%satisfying the cyclic $L_\infty$ algebra by 
%using an alternative method called symmetric construction.
The new gauge-invariant action looks different from the previous one 
but is the same and nothing essentially new.
However, a difference appears when we try to map it to the WZW-like 
action through a field redefinition.
Previously, we could only find a map to the half-WZW-like action \cite{Kunitomo:2019kwk}, 
but the new construction enables us to construct a map to the complete WZW-like action.

\subsection{WZW-like action for the NS-NS sector}

We first summarize the results obtained in \cite{Matsunaga:2016zsu}
on the construction of the WZW-like action for the NS-NS sector. 

For the NS-NS sector, generating functional
of string products with the $L_\infty$ structure, $\bd{L}^{\NSNS}\sst$\,, has been 
constructed by imposing the differential equations \cite{Erler:2014eba},
\begin{subequations}\label{diff eqs EKS} 
 \begin{align}
&\ \partial_t\bd{L}^{\NSNS}\sst\ 
=\ [\bd{L}^{\NSNS}\sst\,, (\bd{\lambda}^{\NSNS}+\bar{\bd{\lambda}}^{\NSNS})\sst]\,,\\
%%%%%%%%%%%%%%%%%
&\ \partial_s\bd{L}^{\NSNS}\sst\ 
=\ [\bd{\eta}\,, \bd{\lambda}^{\NSNS}\sst]\,,\qquad
[\bd{\eta}\,,\bar{\bd{\lambda}}^{\NSNS}\sst]\ =\ 0\,,\\
%%%%%%%%%%%%%%%%%
&\ \partial_{\bar{s}}\bd{L}^{\NSNS}\sst\ 
=\ [\bar{\bd{\eta}}\,, \bar{\bd{\lambda}}^{\NSNS}\sst]\,,\qquad
[\bar{\bd{\eta}}\,,\bd{\lambda}^{\NSNS}\sst]\ =\ 0\,,
\end{align} 
\end{subequations}
introducing the (generating functional of) gauge products $\bd{\lambda}^{\NSNS}\sst$ 
and $\bar{\bd{\lambda}}^{\NSNS}\sst$\,. The type II superstring field theory
with the $L_\infty$ structure is characterized by the $L_\infty$ triplet
$(\bd{\eta},\bar{\bd{\eta}};\bd{L}^{\NSNS})$ through the equations
\begin{subequations}\label{original eqs}
 \begin{align}
 \eta\Phi_{\NSNS}\ =&\ 0\,,
\label{const a}\\
 \bar{\eta}\Phi_{\NSNS}\ =&\ 0\,,
\label{const b}\\
 \pi_1\bd{L}^{\NSNS}(e^{\wedge \Phi_{\NSNS}})\ =&\ 0\,.
\label{eom}
\end{align}
\end{subequations}
The first two (\ref{const a}) and (\ref{const b})
are the constraints imposing $\Phi_{\NSNS}$ is in the small Hilbert space,
and the last one (\ref{eom}) is the equation of motion.
%\textit{dynamical} equation, the equation of motion.
Using the cohomomorphism
\begin{equation}
 \hat{\bd{g}}\ =\ \vec{\mathcal{P}}\exp\left(\int_0^1 
dt (\bd{\lambda}^{\NSNS}+\bar{\bd{\lambda}}^{\NSNS})(0,0,t)\right)\,,
\label{coho g}
\end{equation}
where $\vec{\mathcal{P}}$ denotes the path-ordered product from left to right,
we can transform the original triplet $(\bd{\eta},\bar{\bd{\eta}};\bd{L}^{\NSNS})$
and equations (\ref{original eqs})
to the \textit{dual} $L_\infty$ triplet 
$(\bd{L}^\eta=\hat{\bd{g}}\bd{\eta}\hat{\bd{g}}^{-1},
\bd{L}^{\bar{\eta}}=\hat{\bd{g}}\bar{\bd{\eta}}\hat{\bd{g}}^{-1};\bd{Q})$
and the equations
\begin{subequations}\label{dual eqs}
 \begin{align}
 \pi_1\bd{L}^{\eta}(e^{\wedge \pi_1\hat{\bd{g}}(e^{\wedge \Phi_{\NSNS}})})\ =&\ 0\,,
\label{dual const a}\\
 \pi_1\bd{L}^{\bar{\eta}}(e^{\wedge \pi_1\hat{\bd{g}}(e^{\wedge \Phi_{\NSNS}})})\ =&\ 0\,,
\label{dual const b}\\
 Q\pi_1\hat{\bd{g}}(e^{\wedge \Phi_{\NSNS}})\ =&\ 0\,,
\label{dual eom}
\end{align}
\end{subequations}
respectively, which characterize the WZW-like formulation. 
%The dual triplet $(\bd{L}^\eta, \bd{L}^{\bar{\eta}};\bd{Q})$ 
%is the $L_\infty$ structure characterizing the WZW-like formulation.
The constraint equations, 
\begin{subequations}\label{MC eqs}
 \begin{align}
 \pi_1\bd{L}^{\eta}(e^{\wedge G_{\eta\bar{\eta}}(V)})\ =&\ 0\,,\\
 \pi_1\bd{L}^{\bar{\eta}}(e^{\wedge G_{\eta\bar{\eta}}(V)})\ =&\ 0\,,
\end{align}
\end{subequations}
are identically satisfied by the pure-gauge (functional) string field\footnote{
The method how we find the explicit form of $G_{\eta\bar{\eta}}(V)$ and $B_d(V)$
is given in \cite{Matsunaga:2016zsu}.}
\begin{equation}
 G_{\eta\bar{\eta}}(V)\ =\ \eta\bar{\eta}V + 
\frac{1}{2}\left(L_2^{\eta}(\eta\bar{\eta}V, \bar{\eta}V) + 
\eta L_2^{\bar{\eta}}(\eta\bar{\eta}V, V)\right)  + \cdots\,,
\end{equation}
with the string field $V$ of the WZW-like formulation having the ghost number
$0$ and the picture number $(0,0)$\,. Therefore, if we identify 
\begin{equation}
 G_{\eta\bar{\eta}}(V)\ =\ \pi_1\hat{\bd{g}}(e^{\wedge \Phi_{\NSNS}})\,,
\label{map pure}
\end{equation}
it gives a map between the string fields in two formulations.
By this identification, the last equation (\ref{eom}) becomes 
the equation of motion of the WZW-like formulation:
\begin{equation}
 QG_{\eta\bar{\eta}}(V)\ =\ 0\,.
\label{eom WZW}
\end{equation}
In order to write down the WZW-like action, it is also necessary to introduce
another important functional field called the associated string field, %\footnotemark[11],
\begin{equation}
 B_d(V(t))\ =\ - dV(t) + \frac{1}{2}\left(L^\eta_2(V(t),\bar{\eta}dV(t))
+L^{\bar{\eta}}(\eta V(t),dV(t))\right) + \cdots\,,\quad (d=\partial_t,\delta,Q).
\end{equation}
Here, $V(t)$ is an extension of $V$ by a parameter $t\in[0,1]$ satisfying $V(0)=0$ and $V(1)=V$\,.
The identification (\ref{map pure}) induces the map
\begin{equation}
% B_d(V(t))\ =\ \pi_1\hat{\bd{g}}(\bd{\xi}\bar{\bd{\xi}}\cdot\bd{d})(e^{\wedge \Phi_{\NSNS}})\,,
 B_d(V(t))\ =\ \pi_1\hat{\bd{g}}\bd{D_{\xi\bar{\xi}}}(e^{\wedge \Phi_{\NSNS}})\,,
\label{map asso}
\end{equation}
where %$(\bd{\xi}\bar{\bd{\xi}}\cdot\bd{d})$ 
$\bd{D_{\xi\bar{\xi}}}$ is the coderivation derived from 
%$\pi_1(\bd{\xi}\bar{\bd{\xi}}\cdot\bd{d})=\xi\bar{\xi}d$\,. 
$\pi_1\bd{D_{\xi\bar{\xi}}}=\xi\bar{\xi}\pi_1D$ with 
$D=d$ for $d=\partial_t,\delta$ and $D=\pi_1\bd{L}^{\NSNS}$ for $d=Q$\,.
These maps (\ref{map pure}) and (\ref{map asso}) are consistent with
the identities characterizing the associated field,
\begin{subequations} \label{ident asso}
 \begin{align}
&\ dG_{\eta\bar{\eta}}(V(t))\ =\ D_{\bar{\eta}}(t)D_{\eta}(t)B_d(V(t))\,,
\label{id for asso 1}\\
&\ D_{\bar{\eta}}(t)D_\eta(t)\Big(\partial_tB_\delta(V(t)) - \delta B_{\partial_t}(V(t))\Big)\ =\ 0\,,
\label{id for asso 2}
\end{align}
\end{subequations}
where we introduced the nilpotent linear operators $D_\eta(t)$ and $D_{\bar{\eta}}(t)$ as
\begin{equation}
 D_\eta(t)\varphi\ =\ \pi_1\bd{L}^\eta(e^{\wedge G_{\eta\bar{\eta}}(V(t))}\wedge\varphi)\,,\qquad
 D_{\bar{\eta}}(t)\varphi\ =\ \pi_1\bd{L}^{\bar{\eta}}(e^{\wedge G_{\eta\bar{\eta}}(V(t))}\wedge\varphi)\,,
\end{equation}
for a general string field $\varphi\in\mathcal{H}_l^{\NSNS}$\,. 
Then, we can map the action 
to the WZW-like action
\begin{equation}
 I_{\textrm{WZW}}^{\NSNS}\ =\ \int^1_0 dt\, \omega_l\Big(B_{\partial_t}(V(t)),QG_{\eta\bar{\eta}}(V(t))\Big)\,.
\end{equation}
Using the identities (\ref{ident asso}), 
we can calculate an arbitrary variation of the action as
\begin{equation}
 \delta I^{\NSNS}_{\textrm{WZW}}\ =\ \omega_l\Big(B_\delta(V),QG_{\eta\bar{\eta}}(V)\Big)\,,
\end{equation}
which derives the equation of motion (\ref{eom WZW}) and the gauge invariance 
under the transformation
\begin{equation}
 B_{\delta_g}(V)\ =\ Q\Lambda + D_{\eta}\Omega + D_{\bar{\eta}}\bar{\Omega}\,.
\end{equation}
The gauge transformation generated by $\Lambda$ is mapped from that generated by $\Lambda_{\NSNS}$
in the formulation with the $L_\infty$ structure with the identification
%The identification of the gauge parameters in two formulations are given by
\begin{equation}
 \Lambda\ =\ \pi_1\hat{\bd{g}}\Big(e^{\wedge\Phi_{\NSNS}}\wedge \xi\bar{\xi}\Lambda_{\NSNS}\Big)\,.
\end{equation}
The extra gauge invariances % generated by $\Omega$ and $\bar{\Omega}$ 
come from the fact that the map (\ref{map pure}) is not one-to-one due to the invariance of 
$G_{\eta\bar{\eta}}(V)$ under the variation $B_\delta(V)=D_\eta\Omega+D_{\bar{\eta}}\bar{\Omega}$
following from the identity (\ref{id for asso 1}).

\subsection{Complete WZW-like action and gauge invariance}

It is straightforward to extend the results on the NS-NS sector to all four sectors. 
We first note that the similarity transformation by $\hat{\bd{F}}$ is trivial
except for the NS-NS sector, and thus, the string product $\bd{L}$ we constructed reduces to
%$\bd{L}|^{0,0}$\,,
\begin{equation}
 \bd{L}|^{(0,0)}\ =\ 
%\bd{Q}+\bd{B}|^{(0,0)}\,,\qquad
\bd{Q}+ \sum_{p=0}^\infty \bd{B}^{(p,p)}_{p+1}|^{(0,0)}\ \equiv\
\bd{Q}+\bd{B}|^{(0,0)}\,,
\end{equation}
in the NS-NS sector.
%$\bd{L}=\bd{Q}+\bd{B}|^{(0,0)}$
%$L_\infty$ triplet becomes $(\bd{\eta},\bar{\bd{\eta}};\bd{L}=\bd{Q}+\bd{B})$ in the NS-NS sector. 
Then, the differential equations (\ref{diff eqs}) implies that
the generating functional $\bd{L}|^{(0,0)}\sst=\bd{Q}+\bd{B}|^{(0,0)}\sst$ of 
string product in the NS-NS sector satisfies
\begin{subequations}\label{diff eqs NSNS}
\begin{align}
&\ \partial_t\bd{L}|^{(0,0)}\sst\ 
=\ [\,\bd{L}|^{(0,0)}\sst\,, (\bd{\lambda}+\bar{\bd{\lambda}})|^{(0,0)}\sst\,]\,,\\
%%%%%%%%%%%%%%%%%
&\ \partial_s\bd{L}|^{(0,0)}\sst\ 
=\ [\,\bd{\eta}\,, \bd{\lambda}|^{(0,0)}\sst\,]\,,\qquad
[\,\bd{\eta}\,,\bar{\bd{\lambda}}|^{(0,0)}\sst\,]\ =\ 0\,,\\
%%%%%%%%%%%%%%%%%
&\ \partial_{\bar{s}}\bd{L}|^{(0,0)}\sst\ 
=\ [\,\bar{\bd{\eta}}\,, \bar{\bd{\lambda}}|^{(0,0)}\sst\,]\,,\qquad
[\,\bar{\bd{\eta}}\,,\bd{\lambda}|^{(0,0)}\sst\,]\ =\ 0\,,
\end{align}
\end{subequations}
where
\begin{subequations}
 \begin{align}
 \bd{B}|^{(0,0)}\sst\ =&\ \sum_{p,\bar{p},m,\bar{m}=0}^\infty
%\sum_{\bar{p},\bar{m}=0}^\infty 
t^{p+\bar{p}}s^m \bar{s}^{\bar{m}}
\bd{B}^{(p,\bar{p})}_{p+m+1,\bar{p}+\bar{m}+1}|^{(0,0)}\,,\\
%=\ \sum_{m,\bar{m}=0}^\infty s^m\bar{s}^{\bar{m}}\bd{B}^{[m,\bar{m}]}(t)|^{(0,0)}\,,\\
%%%%%%%%%%%%%%%%%%%%%%%%
 \bd{\lambda}|^{(0,0)}\sst\ =&\ \sum_{p,\bar{p},m,\bar{m}=0}^\infty
%\sum_{p,m=0}^\infty\sum_{\bar{p},\bar{m}=0}^\infty 
t^{p+\bar{p}}s^m \bar{s}^{\bar{m}}
\bd{\lambda}^{(p+1,\bar{p})}_{p+m+2,\bar{p}+\bar{m}+1}|^{(0,0)}\,,\\
%%%%%%%%%%%%%%%%%%%%%%%%
 \bar{\bd{\lambda}}|^{(0,0)}\sst\ =&\ \sum_{p,\bar{p},m,\bar{m}=0}^\infty
%\sum_{p,m=0}^\infty\sum_{\bar{p},\bar{m}=0}^\infty 
t^{p+\bar{p}}s^m \bar{s}^{\bar{m}}
\bar{\bd{\lambda}}^{(p,\bar{p}+1)}_{p+m+1,\bar{p}+\bar{m}+2}|^{(0,0)}\,.
\end{align}
\end{subequations}
Since the equations (\ref{diff eqs NSNS}) agrees with the equations
(\ref{diff eqs EKS}), we can identify the new string product in the NS-NS sector,
$\bd{L}|^{(0,0)}=\bd{L}|^{(0,0)}(0,0,1)$\,,
%\begin{equation}
% \bd{L}(0,0,1)|^{(0,0)}\equiv\ \bd{L}^{\NSNS}\ =\ 
%\bd{Q} + \bd{B}|^{(0,0)}\,,\qquad
%\bd{B}|^{(0,0)}\ =\ \sum_{n=1}^\infty\bd{B}_{n+1}^{(n,n)}|^{(0,0)}\,,
%\end{equation}
to that constructed in \cite{Erler:2014eba}. The cohomomorphism 
(\ref{coho g}) is then given by
\begin{equation}
 \hat{\bd{g}}\ =\ \vec{\mathcal{P}}\exp\left(\int_0^1 
dt (\bd{\lambda}+\bar{\bd{\lambda}})|^{(0,0)}(0,0,t)\right)\,.
\end{equation}
%The original triplet $(\bd{\eta},\bar{\bd{\eta}};\bd{L}^{\NSNS})$ and the dual triplet
%$(\bd{L}^{\eta},\bd{L}^{\bar{\eta}};\bd{Q})$ are 
With this identification, the $L_\infty$ triplet $(\bd{\eta},\bar{\bd{\eta}};\bd{L})$ 
can transform to the dual triplet $(\bd{L}^\eta, \bd{L}^{\bar{\eta}};\tilde{\bd{L}})$\,,
where
\begin{align}
 \pi_1\tilde{\bd{L}}\ =&\ \pi_1\hat{\bd{g}}\bd{L}\hat{\bd{g}}^{-1}\ 
=\ \pi_1\bd{Q} + \mathcal{G}\pi_1\tilde{\bd{b}}\,,\\
\pi_1\tilde{\bd{b}}\ =&\ \pi_1\hat{\bd{g}}(\bd{b}-\bd{B}|^{(0,0)})\hat{\bd{g}}^{-1}\,.
\end{align}
%with
%\begin{equation}
%\pi_1\tilde{\bd{b}}\ =\ \pi_1\hat{\bd{g}}(\bd{b}-\bd{B}|^{(0,0)})\hat{\bd{g}}^{-1}\,.
%%\pi_1\bd{b}\ =\ \pi_1\bd{B}\hat{\bd{F}}\,.
%\end{equation}
Since $\hat{\bd{g}}$ is identity except in the NS-NS sector, we find that
\begin{equation}
\pi_1\hat{\bd{g}}(e^{\wedge\Phi})\ =\
\pi_1\hat{\bd{g}}(e^{\wedge\Phi_{\NSNS}}) + \Phi_{\RNS} + \Phi_{\NSR} + \Phi_{\RR} \,,			
\end{equation}
and can identify $\Phi_{\RNS}$\,, $\Phi_{\NSR}$\,, and $\Phi_{\RR}$ with
the corresponding string fields $\Psi$\,, $\bar{\Psi}$\,, and $\Sigma$\,,
in the WZW-like formulation, respectively:
%That is, if we denote them $\Psi$\,, $\bar{\Psi}$\,, and $\Sigma$\,, respectively, we have
\begin{equation}
\Phi_{\RNS}\ =\ \Psi\,,\quad \Phi_{\RNS}\ =\ \bar{\Psi}\,,\quad \Phi_{\RR}\ =\ \Sigma\,.
\label{id others}
\end{equation}
%where we denoted the string fields in the WZW-like formulation as
%of the WZW-like formulation, 
%to those in the formulation with $L_\infty$ structure as
%we have
%\begin{equation}
%\Phi_{\RNS}\ =\ \Psi\,,\quad \Phi_{\RNS}\ =\ \bar{\Psi}\,,\quad \Phi_{\RR}\ =\ \Sigma\,.
%\label{id others}
%\end{equation}
The identification (\ref{map pure}) is now extended as
\begin{equation}
\pi_1\hat{\bd{g}}(e^{\wedge\Phi})  
=\ \ G_{\eta\bar{\eta}}(V) + \Psi + \bar{\Psi} + \Sigma\
\equiv\ G(\mathcal{V})\,,
\end{equation}
where we denoted %the fields %V,\, \Psi,\, \bar{\Psi},\, $ and $\Sigma$
the string fields in the WZW-formulation as $\mathcal{V}$ collectively. 
The dual constraint equations 
(\ref{dual const a}) and (\ref{dual const b}) map to
\begin{align}
 0\ =&\ 
%\pi_1\hat{\bd{g}}\bd{\eta}(e^{\wedge \Phi})\ 
\pi_1\bd{L}^\eta(e^{\wedge G(\mathcal{V})})\
=\ \pi_1\bd{L}^\eta(e^{\wedge G_{\eta\bar{\eta}}(V)}) 
+ \eta\Psi + \eta\bar{\Psi} + \eta\Sigma\,,\\
%%%%%%%%%%%%%%%%%%%%%%%%%%
 0\ =&\ 
%\pi_1\hat{\bd{g}}\bar{\bd{\eta}}(e^{\wedge \Phi})
\pi_1\bd{L}^{\bar{\eta}}(e^{\wedge G(\mathcal{V})})
=\ \pi_1\bd{L}^{\bar{\eta}}(e^{\wedge G_{\eta\bar{\eta}}(V)}) 
+ \bar{\eta}\Psi + \bar{\eta}\bar{\Psi} + \bar{\eta}\Sigma\,,
\end{align}
which decompose
to the Maurer-Cartan equations (\ref{MC eqs}) 
%for the pure-gauge-like string field $G_{\eta\bar{\eta}}(V)$
and the constraints that $\Psi$\,, $\bar{\Psi}$\,, and $\Sigma$ are the fields
in the small Hilbert space. The map for the associated string field (\ref{map asso})
can also be extended to that for all four sectors as 
%$\mathcal{B}_d(\mathcal{V}(t))$\,, as
%\begin{equation}
% \mathcal{B}_d(\mathcal{V}(t))\ =\ \pi_1\hat{\bd{g}}(\bd{\xi}\bar{\bd{\xi}}\cdot\bd{d})(e^{\wedge\Phi})\,,
%\end{equation}
%which can be written by the identificaitons
%(\ref{id NSNS}) and (\ref{id others}) as
%\begin{equation}
%  \mathcal{B}_d(\mathcal{V}(t))\ =\ 
%B_d(V(t)) + \xi\bar{\xi}d\Psi(t)
%+ \xi\bar{\xi}d\bar{\Psi}(t)
%+ \xi\bar{\xi}d\Sigma(t)\,.
%\end{equation}
%
\begin{equation}
\pi_1\hat{\bd{g}}%(\bd{\xi}\bar{\bd{\xi}}\cdot\bd{d})
\bd{D_{\xi{\bar{\xi}}}}(e^{\wedge \Phi})\ =\
B_d(V(t)) + \xi\bar{\xi}d\Psi(t)
+ \xi\bar{\xi}d\bar{\Psi}(t)
+ \xi\bar{\xi}d\Sigma(t)\
\equiv \mathcal{B}_d(\mathcal{V}(t))\,. 
\end{equation}
We can find that the identities (\ref{ident asso}) are then extended to
\begin{align}
&\ d G(\mathcal{V}(t))\ =\ D_{\bar{\eta}}(t)D_\eta(t)\mathcal{B}_d(\mathcal{V}(t))\,,\\
&\ D_{\bar{\eta}}(t)D_\eta(t)\Big(\partial_t\mathcal{B}_\delta(\mathcal{V}(t)) 
- \delta\mathcal{B}_\delta(\mathcal{V}(t))\Big)\ =\ 0\,.
\end{align}
Here, for general string field $\varphi\in\mathcal{H}$\,,
\begin{align}
 D_\eta(t)\varphi\ =&\ 
\pi_1^{(0,0)}\bd{L}^\eta(e^{\wedge G_{\eta\bar{\eta}}(V(t))}\wedge\varphi_{\NSNS})
+ \eta\varphi_{\RNS} + \eta\varphi_{\RNS} + \eta\varphi_{\RR}\,,\\
%%%%%%%%%%%%%%%%%%%%%%%%%%%%%
 D_{\bar{\eta}}(t)\varphi\ =&\ 
\pi_1^{(0,0)}\bd{L}^{\bar{\eta}}(e^{\wedge G_{\eta\bar{\eta}}(V(t))}\wedge\varphi_{\NSNS})
+ \bar{\eta}\varphi_{\RNS} + \bar{\eta}\varphi_{\NSR} + \bar{\eta}\varphi_{\RR}\,,
\end{align}
with 
%$\varphi_{\NSNS}=\pi^{(0,0)}\varphi\,,\
%\varphi_{\RNS}=\pi^{(1,0)}\varphi\,,\
%\varphi_{\NSR}=\pi^{(0,1)}\varphi\,,\
%\varphi_{\RR}=\pi^{(1,1)}\varphi$\,.
\begin{equation}
\varphi_{\NSNS}\ =\ \pi^{(0,0)}\varphi\,,\quad
\varphi_{\RNS}\ =\ \pi^{(1,0)}\varphi\,,\quad
\varphi_{\NSR}\ =\ \pi^{(0,1)}\varphi\,,\quad
\varphi_{\RR}\ =\ \pi^{(1,1)}\varphi\,.
\end{equation}
It is easy to show that an arbitrary variation of the action
\begin{equation}
 I_{\textrm{WZW}}\ =\ \int^1_0 dt\, \omega_l\Big(\mathcal{B}_{\partial_t}(\mathcal{V}(t))\,, 
\mathcal{G}^{-1}\pi_1\tilde{\bd{L}}(e^{\wedge G(\mathcal{V}(t))})\Big)
\end{equation}
gives
\begin{equation}
 \delta I_{\textrm{WZW}}\ =\ 
\omega_l\Big(\mathcal{B}_\delta(\mathcal{V}),\mathcal{G}^{-1}\pi_1\tilde{\bd{L}}
(e^{\wedge G(\mathcal{V})})\Big)\,.
\end{equation}
%which gives 
It derives the equations of motion
\begin{equation}
 \tilde{\bd{L}}(e^{\wedge G(\mathcal{V})})\ =\ 0
\end{equation}
and yields the invariance of the action under the gauge transformation
\begin{equation}
 \mathcal{B}_{\delta_{\Lambda}}(\mathcal{V})\ =\ \pi_1^{(0,0)}\tilde{\bd{L}}(e^{\wedge G}\wedge\hat{\Lambda})
+ \xi\bar{\xi}(\pi_1^{(1,0)}+\pi_1^{(0,1)}+\pi_1^{(1,1)})
\tilde{\bd{L}}(e^{\wedge G}\wedge \bar{\eta}\eta\hat{\Lambda})\,,
\end{equation}
%
%\begin{align}
%B_{\delta_{\Lambda}}\ =&\ \pi_1^{(0,0)}\tilde{\bd{L}}(e^{\wedge G}\wedge\hat{\Lambda})\,,\\
%%%%%%%%%%%%%%%%%%%
%\delta_{\Lambda}\Psi\ =&\ 
%Q\lambda + X\pi_1^{(1,0)}\tilde{\bd{b}}(e^{\wedge G}\wedge\bar{\eta}\eta\hat{\Lambda})\,,\\
%%%%%%%%%%%%%%%%%%%
%\delta_{\Lambda}\bar{\Psi}\ =&\
%Q\bar{\lambda} + \bar{X}\pi_1^{(0,1)}\tilde{\bd{b}}(e^{\wedge G}\wedge\bar{\eta}\eta\hat{\Lambda})\,,\\
%%%%%%%%%%%%%%%%%%%
%\delta_{\Lambda}\Sigma\ =&\
%Q\rho + X\bar{X}\pi_1^{(1,1)}\tilde{\bd{b}}(e^{\wedge G}\wedge\bar{\eta}\eta\hat{\Lambda})\,,\\
%\end{align}
%
where 
\begin{equation}
\pi_1^{(0,0)}\hat{\Lambda}=\Lambda\,,\quad
\pi_1^{(1,0)}\hat{\Lambda}=\xi\bar{\xi}\lambda\,,\quad
\pi_1^{(0,0)}\hat{\Lambda}=\xi\bar{\xi}\bar{\lambda}\,, \quad
\pi_1^{(0,0)}\hat{\Lambda}=\xi\bar{\xi}\rho\,,  
\end{equation}
%\begin{equation}
% \hat{\Lambda}\ =\ \Lambda+\xi\bar{\xi}\lambda+\xi\bar{\xi}\bar{\lambda}+\xi\bar{\xi}\rho
%\end{equation}
with
$\eta\lambda=\bar{\eta}\lambda=\eta\bar{\lambda}=\bar{\eta}\bar{\lambda}=\eta\rho=\bar{\eta}\rho=0$\,.
The identification with the parameters of the formulation with $L_\infty$ structure is given by
\begin{equation}
 \tilde{\Lambda}\ =\ \pi_1^{(0,0)}\hat{\bd{g}}(e^{\wedge \Phi_{\NSNS}}\wedge \xi\bar{\xi}\Lambda_{\NSNS})\,,
\quad \lambda\ =\ \Lambda_{\RNS}\,,\quad \bar{\lambda}\ =\ \Lambda_{\NSR}\,,\quad \rho\ =\ \Lambda_{\RR}\,.
\end{equation}

\sectiono{Conclusion and discussion}\label{concl}

In this paper, we have revisited the type II superstring field theory 
with $L_\infty$ structure and proposed an alternative 
method to construct string products, which is symmetric with respect to the left- and 
right-moving sectors. The symmetric method makes transparent not only the construction of 
string products but also the proof that the tree-level S-matrix agrees with that calculated 
using the first-quantization method.
Another advantage of the symmetric construction is that it enables us to write down a WZW-like 
action through a map between the string fields of the two formulations, which was not possible 
with the previous (asymmetric) construction method. 
%We have written down a WZW-like action by giving a map explicitly. 
The complete WZW-like action of type II superstring field theory, 
which was the only missing piece, has now been constructed.
We have completed the superstring field theory for all three complementary formulations, 
which allows us to use a convenient formulation depending on what we are studying.

There is another interesting superstring field theory to consider: the (oriented) open-closed 
superstring field theory. This is the one that should be derived as type II superstring field 
theory on a non-trivial D-brane background but is worth constructing independently as it helps 
to calculate non-perturbative effects. In fact, the one based on the formulation with extra
free field has already been constructed \cite{Moosavian:2019ydz}
and used for studying some non-perturbative effect 
\cite{Sen:2019qqg,Sen:2020cef,Sen:2020oqr,Sen:2020ruy,Sen:2021qdk}.
It is interesting to construct it based on the open-closed homotopy algebra (OCHA) \cite{Kajiura:2004xu} 
and study the relation to the one in the WZW-like formulation.

\vspace{1cm}
\noindent
{\bf \large Acknowledgments}

This work is supported in part by JSPS Grant-in-Aid for Scientific 
Research (C) Grant Number JP18K03645.

\vspace{1cm}
\appendix

\sectiono{Projected commutators}\label{pro comm}

The definition of the projected commutator introduced in (\ref{L infinity B}) 
is given by
\begin{subequations}\label{proj comm}  
\begin{align}
[\bd{\mathfrak{D}}\,, \bd{\mathfrak{D}}']^{11}\ =&\
\sum_{n,r,\bar{r}}\sum_{m,s,\bar{s}}
 [\,\bd{\mathfrak{D}}_n|^{(2r,2\bar{r})}\,, 
  \bd{\mathfrak{D}}'_m|^{(2s,2\bar{s})}]\mid^{(2(r+s),2(\bar{r}+\bar{s}))}\,,\\
%%%%%
[\bd{\mathfrak{D}}\,, \bd{\mathfrak{D}}']^{21}\ =&\
\sum_{n,r,\bar{r}}\sum_{m,s,\bar{s}}
 [\,\bd{\mathfrak{D}}_n|^{(2r,2\bar{r})}\,, 
\bd{\mathfrak{D}}'_m|^{(2s,2\bar{s})}]\mid^{(2(r+s-1),2(\bar{r}+\bar{s}))}\,,\\
%%%%%
[\bd{\mathfrak{D}}\,, \bd{\mathfrak{D}}']^{12}\ =&\
\sum_{n,r,\bar{r}}\sum_{m,s,\bar{s}}
 [\,\bd{\mathfrak{D}}_n|^{(2r,2\bar{r})}\,, 
\bd{\mathfrak{D}}'_m|^{(2s,2\bar{s})}]\mid^{(2(r+s),2(\bar{r}+\bar{s}-1))}\,,\\
%%%%%
[\bd{\mathfrak{D}}\,, \bd{\mathfrak{D}}']^{22}\ =&\
\sum_{n,r,\bar{r}}\sum_{m,s,\bar{s}}
 [\,\bd{\mathfrak{D}}_n|^{(2r,2\bar{r})}\,, 
\bd{\mathfrak{D}}_m|^{(2s,2\bar{s})}]\mid^{(2(r+s-1),2(\bar{r}+\bar{s}-1))}\,,
\end{align}
\end{subequations}
for coderivations 
$\bd{\mathfrak{D}}=\sum_{n,r,\bar{r}}\bd{\mathfrak{D}}_n|^{(2r,2\bar{r})}$ 
and $\bd{\mathfrak{D}}'=\sum_{m,s,\bar{s}}\bd{\mathfrak{D}}'_m|^{(2s,2\bar{s})}$\,.
An alternative expressions obtained by projecting the intermediate state \cite{Kunitomo:2020xrl},
\begin{subequations}
 \begin{align}
[\,\bd{\mathfrak{D}}\,, \bd{\mathfrak{D}}']^{11}\ =&\
\sum_n\left(\bd{\mathfrak{D}}_n\big(\pi_1^{(0,0)}\bd{\mathfrak{D}}'\wedge\id_{n-1}\big) 
- (-)^{|\mathfrak{D}||\mathfrak{D}'|}\bd{\mathfrak{D}}'_n\big(\pi_1^{(0,0)}\bd{\mathfrak{D}}
\wedge\id_{n-1}\big)\right)\,,\\
%%%%%%%%%%%%%%%
[\,\bd{\mathfrak{D}}\,, \bd{\mathfrak{D}}']^{21}\ =&\
\sum_n\left(\bd{\mathfrak{D}}_n\big(\pi_1^{(1,0)}\bd{\mathfrak{D}}'\wedge\id_{n-1}\big) 
- (-)^{|\mathfrak{D}||\mathfrak{D}'|}\bd{\mathfrak{D}}'_n\big(\pi_1^{(1,0)}\bd{\mathfrak{D}}
\wedge\id_{n-1}\big)\right)\,,\\
%%%%%%%%%%%%%%%
[\,\bd{\mathfrak{D}}\,, \bd{\mathfrak{D}}']^{12}\ =&\
\sum_n\left(\bd{\mathfrak{D}}_n\big(\pi_1^{(0,1)}\bd{\mathfrak{D}}'\wedge\id_{n-1}\big) 
- (-)^{|\mathfrak{D}||\mathfrak{D}'|}\bd{\mathfrak{D}}'_n\big(\pi_1^{(0,1)}\bd{\mathfrak{D}}
\wedge\id_{n-1}\big)\right)\,,\\
%%%%%%%%%%%%%%%
[\,\bd{\mathfrak{D}}\,, \bd{\mathfrak{D}}']^{22}\ =&\
\sum_n\left(\bd{\mathfrak{D}}_n\big(\pi_1^{(1,1)}\bd{\mathfrak{D}}'\wedge\id_{n-1}\big) 
- (-)^{|\mathfrak{D}||\mathfrak{D}'|}\bd{\mathfrak{D}}'_n\big(\pi_1^{(1,1)}\bd{\mathfrak{D}}
\wedge\id_{n-1}\big)\right)\,,
 \end{align}
\end{subequations}
are also useful. Here, $|\mathfrak{D}|$ ($|\mathfrak{D}'|$) is equal to 0 or 1 when the degree of coderivation
$\bd{\mathfrak{D}}$ ($\bd{\mathfrak{D}}'$) is even or odd, respectively.
The square bracket with subscript $\mathfrak{X}=X$ or $\bar{X}$ is defined by inserting $\mathfrak{X}$
at the intermediate state\footnote{
These brackets can only be applicable for the case considered
since $X$ and $\bar{X}$ are the PCO's acting on the states with picture number $-3/2$\,.}:
\begin{equation}
[\,\bd{\mathfrak{D}}\,, \bd{\mathfrak{D}}']^{22}_{\mathfrak{X}}\ =\
\sum_n\left(\bd{\mathfrak{D}}_n\big(\mathfrak{X}\pi_1^{(1,1)}\bd{\mathfrak{D}}'\wedge\id_{n-1}\big) 
- (-)^{|\mathfrak{D}||\mathfrak{D}'|}\bd{\mathfrak{D}}'_n
\big(\mathfrak{X}\pi_1^{(1,1)}\bd{\mathfrak{D}}\wedge\id_{n-1}\big)\right)\,.\\
\end{equation}
Similarly, the square bracket with subscript $\mathcal{O}=\Xi$ or $\bar{\Xi}$ is defined by 
\begin{equation}
 [\bd{\mathfrak{D}}\,,\bd{\mathfrak{D}}']^{22}_{\mathcal{O}}\ =\ 
\sum_n\left(\bd{\mathfrak{D}}_n(\mathcal{O}\pi_1^{(1,1)}\bd{\mathfrak{D}}'\wedge \id^{\wedge (n-1)})
 + (-1)^{(|\mathfrak{D}|+1)(|\mathcal{D}'|+1)}
\bd{\mathfrak{D}}'_n(\mathcal{O}\pi_1^{(1,1)}\bd{\mathfrak{D}}\wedge \id^{\wedge (n-1)})\right)\,.
\end{equation}
The projected commutators (\ref{proj comm}) satisfy the Jacobi identities
\begin{equation}
 [\,\bd{\mathfrak{D}}\,, [\,\bd{\mathfrak{D}}'\,, \bd{\mathfrak{D}}''\,]^{ab}]^{cd}
+  [\,\bd{\mathfrak{D}}\,, [\,\bd{\mathfrak{D}}'\,, \bd{\mathfrak{D}}''\,]^{cd}]^{ab}
+ (\textrm{cyclic}\ \textrm{perm.})\ =\ 0\,,
\end{equation}
which reduce to the conventional ones if two projected commutators are the same type.
Similar identities also hold even if they include projected commutator(s) 
with subscript $\mathfrak{X}=X$ or $\bar{X}$\,:
\begin{align}
&\ [\,\bd{\mathfrak{D}}\,, [\,\bd{\mathfrak{D}}'\,, \bd{\mathfrak{D}}''\,]^{22}_{\mathfrak{X}}]^{ab}
+  [\,\bd{\mathfrak{D}}\,, [\,\bd{\mathfrak{D}}'\,, \bd{\mathfrak{D}}''\,]^{ab}]^{22}_{\mathfrak{X}}
+ (\textrm{cyclic}\ \textrm{perm.})\ =\ 0\,,\\
%%%%%%%%%%%%%%%%%%
&\ [\,\bd{\mathfrak{D}}\,, [\,\bd{\mathfrak{D}}'\,, \bd{\mathfrak{D}}''\,]^{22}_{\mathfrak{X}}]^{22}_{\mathfrak{X}'}
+  [\,\bd{\mathfrak{D}}\,, [\,\bd{\mathfrak{D}}'\,, \bd{\mathfrak{D}}''\,]^{22}_{\mathfrak{X}'}]^{22}_{\mathfrak{X}}
+ (\textrm{cyclic}\ \textrm{perm.})\ =\ 0\,.
\end{align}
The additional sign factor is necessary for the Jacobi identities including the one with 
subscript $\mathcal{O}=\Xi$ or $\bar{\Xi}$ since $\Xi$ and $\bar{\Xi}$ are the Grassmann odd operators. 
For example,
\begin{align}
&\ [\,\bd{\mathfrak{D}}\,, [\,\bd{\mathfrak{D}}'\,, \bd{\mathfrak{D}}''\,]^{22}_{\mathcal{O}}]^{ab}
+ (-1)^{|\mathfrak{D}'|} [\,\bd{\mathfrak{D}}\,, [\,\bd{\mathfrak{D}}'\,, 
\bd{\mathfrak{D}}''\,]^{ab}]^{22}_{\mathcal{O}}
+ (\textrm{cyclic}\ \textrm{perm.})\ =\ 0\,,\\
%%%%%%%%%%%%%%%%%%%%%%%%%
&\ [\,\bd{\mathfrak{D}}\,, [\,\bd{\mathfrak{D}}'\,, \bd{\mathfrak{D}}''\,]^{22}_{\mathcal{O}}]^{22}_{\mathfrak{X}}
+ (-1)^{|\mathfrak{D}'|} [\,\bd{\mathfrak{D}}\,, 
[\,\bd{\mathfrak{D}}'\,, \bd{\mathfrak{D}}''\,]^{22}_{\mathfrak{X}}]^{22}_{\mathcal{O}}
+ (\textrm{cyclic}\ \textrm{perm.})\ =\ 0\,.
\end{align}
%for coderivations equal to neither $\bd{Q}$\,, $\bd{\eta}$\,, nor $\bar{\bd{\eta}}$\,.
%If one of the coderivations is equal to $\bd{Q}$\,, $\bd{\eta}$\,, or $\bar{\bd{\eta}}$\,,
%there is an additional contribution if they do not commute with $\mathcal{O}$\,, for example,
We should note that $\mathcal{O}$ is not commutative to $Q$ and either $\eta$ or $\bar{\eta}$\,.
Thus, for example, 
\begin{align}
%%%%%%%%%%%%%%%%%
 [\,\bd{Q}\,,[\,\bd{\mathfrak{D}}\,, \bd{\mathfrak{D}'}\,]^{22}_\Xi\,]\ =&\
[\,[\,\bd{Q}\,, \bd{\mathfrak{D}}\,]\,, \bd{\mathfrak{D}}'\,]^{22}_\Xi
+ (-1)^{|\mathfrak{D}|+1}[\,\bd{\mathfrak{D}}\,, [\,\bd{Q}\,, \bd{\mathfrak{D}}'\,]\,]^{22}_\Xi
+ (-1)^{|\mathfrak{D}|}[\,\bd{\mathfrak{D}}\,, \bd{\mathfrak{D}}'\,]^{22}_X\,,\\
%%%%%%%%%%%%%%%%%
% [\,\bd{Q}\,,[\,\bd{\mathfrak{D}}\,, \bd{\mathfrak{D}'}\,]^{22}_{\bar{\Xi}}\,]\ =&\
%[\,[\,\bd{Q}\,, \bd{\mathfrak{D}}\,]\,, \bd{\mathfrak{D}}'\,]^{22}_{\bar{\Xi}}
%+ (-1)^{|\mathfrak{D}|+1}[\,\bd{\mathfrak{D}}\,, [\,\bd{Q}\,, \bd{\mathfrak{D}}'\,]\,]^{22}_{\bar{\Xi}}
%+ (-1)^{|\mathfrak{D}|}[\,\bd{\mathfrak{D}}\,, \bd{\mathfrak{D}}'\,]^{22}_{\bar{X}}\,,\\
%%%%%%%%%%%%%%%%%
 [\,\bd{\eta}\,,[\,\bd{\mathfrak{D}}\,, \bd{\mathfrak{D}'}\,]^{22}_\Xi\,]\ =&\
[\,[\,\bd{\eta}\,, \bd{\mathfrak{D}}\,]\,, \bd{\mathfrak{D}}'\,]^{22}_\Xi
+ (-1)^{|\mathfrak{D}|+1}[\,\bd{\mathfrak{D}}\,, [\,\bd{\eta}\,, \bd{\mathfrak{D}}'\,]\,]^{22}_\Xi
+ (-1)^{|\mathfrak{D}|}[\,\bd{\mathfrak{D}}\,, \bd{\mathfrak{D}}'\,]^{22}\,.%\\
%%%%%%%%%%%%%%%%%
% [\,\bd{\eta}\,,[\,\bd{\mathfrak{D}}\,, \bd{\mathfrak{D}'}\,]^{22}_{\bar{\Xi}}\,]\ =&\
%[\,[\,\bd{\eta}\,, \bd{\mathfrak{D}}\,]\,, \bd{\mathfrak{D}}'\,]^{22}_{\bar{\Xi}}
%+ (-1)^{|\mathfrak{D}|+1}[\,\bd{\mathfrak{D}}\,, [\,\bd{\eta}\,, \bd{\mathfrak{D}}'\,]\,]^{22}_{\bar{\Xi}}\,,\\
%%%%%%%%%%%%%%%%%
% [\,\bar{\bd{\eta}}\,,[\,\bd{\mathfrak{D}}\,, \bd{\mathfrak{D}'}\,]^{22}_\Xi\,]\ =&\
%[\,[\,\bar{\bd{\eta}}\,, \bd{\mathfrak{D}}\,]\,, \bd{\mathfrak{D}}'\,]^{22}_\Xi
%+ (-1)^{|\mathfrak{D}|+1}[\,\bd{\mathfrak{D}}\,, [\,\bar{\bd{\eta}}\,, \bd{\mathfrak{D}}'\,]\,]^{22}_\Xi\,,\\
%%%%%%%%%%%%%%%%%
% [\,\bar{\bd{\eta}}\,,[\,\bd{\mathfrak{D}}\,, \bd{\mathfrak{D}'}\,]^{22}_{\bar{\Xi}}\,]\ =&\
%[\,[\,\bar{\bd{\eta}}\,, \bd{\mathfrak{D}}\,]\,, \bd{\mathfrak{D}}'\,]^{22}_{\bar{\Xi}}
%+ (-1)^{|\mathfrak{D}|+1}[\,\bd{\mathfrak{D}}\,, [\,\bar{\bd{\eta}}\,, \bd{\mathfrak{D}}'\,]\,]^{22}_{\bar{\Xi}}
%+ (-1)^{|\mathfrak{D}|}[\,\bd{\mathfrak{D}}\,, \bd{\mathfrak{D}}'\,]^{22}\,,
\end{align}

\sectiono{Expansion of $\bd{B}_3^{(\bullet,\bullet)}\ssb$, 
$\bd{\lambda}_3^{(\bullet,\bullet)}\ssb$ and 
$\bar{\bd{\lambda}}_3^{(\bullet,\bullet)}\ssb$}
\label{explicit B3}

\begin{figure}[hbtp]
  \begin{center}
   \includegraphics[clip,width=16cm]{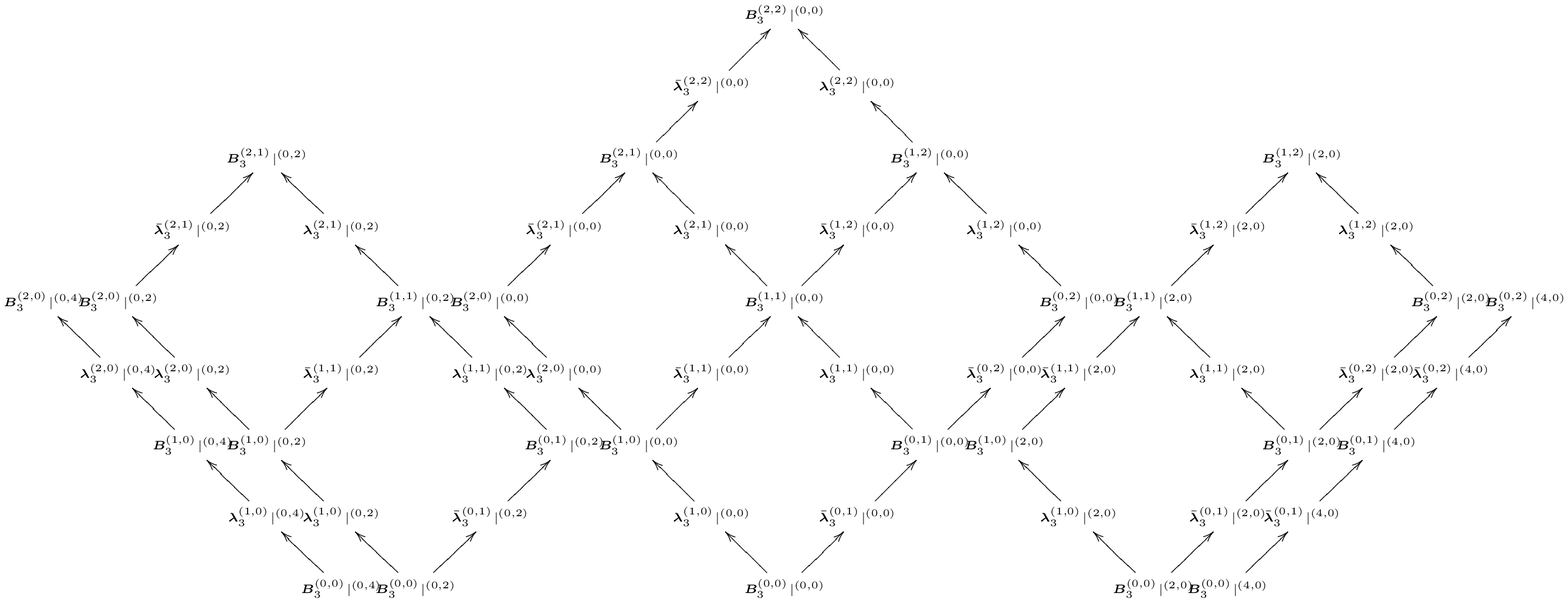}
\caption{The flow of how each component of 3-string products are determined 1.}
\label{3string comp1}
  \end{center}
\end{figure}
\begin{figure}[hbtp]
  \begin{center}
   \includegraphics[clip,width=10cm]{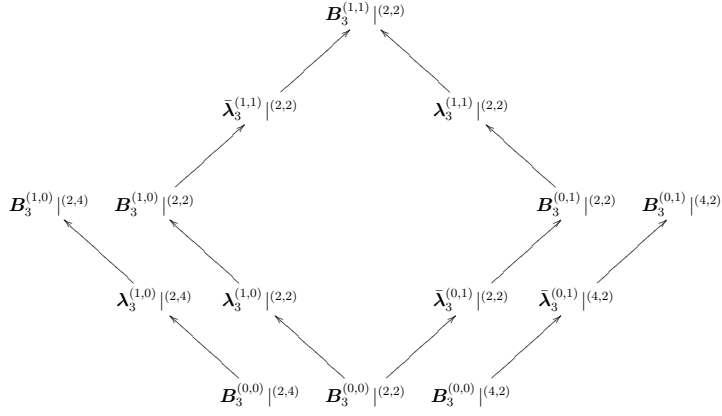}
\caption{The flow of how each component of 3-string products are determined 2.}
\label{3string comp2}
  \end{center}
\end{figure}

In section \ref{sym const}, we show that how the generating functionals of 
3-string and gauge products with specific picture number are determined.
In this appendix, we illustrate the flow to be determined the 3-string (gauge) products 
with specific cyclic Ramond numbers, which are given as coefficients 
further expanded by $s$ and $\bar{s}$\,:
\begin{align}
 \bd{B}_3^{(0,0)}\ssb\ =&\ \bd{B}_3^{(0,0)}\mid^{(4,4)} + s \bd{B}_3^{(0,0)}\mid^{(2,4)}
+ \bar{s} \bd{B}_3^{(0,0)}\mid^{(4,2)} 
\nonumber\\
&\
+ s^2 \bd{B}_3^{(0,0)}\mid^{(0,4)}
+ s\bar{s} \bd{B}_3^{(0,0)}\mid^{(2,2)} + \bar{s}^2 \bd{B}_3^{(0,0)}\mid^{(4,0)}
\nonumber\\
&\
+ s^2\bar{s} \bd{B}_3^{(0,0)}\mid^{(0,2)} + s\bar{s}^2 \bd{B}_3^{(0,0)}\mid^{(2,0)}
+ s^2\bar{s}^2 \bd{B}_3^{(0,0)}\mid^{(0,0)}\,,\\
%%%%%%%%%%%%%%%%%%%%%%%%
\bd{B}_3^{(1,0)}\ssb\ =&\ \bd{B}_3^{(1,0)}\mid^{(2,4)} + s \bd{B}_3^{(1,0)}\mid^{(0,4)}
+ \bar{s} \bd{B}_3^{(1,0)}\mid^{(2,2)} 
\nonumber\\
&\
+ s\bar{s} \bd{B}_3^{(1,0)}\mid^{(0,2)}
+ \bar{s}^2 \bd{B}_3^{(1,0)}\mid^{(2,0)} + s\bar{s}^2 \bd{B}_3^{(1,0)}\mid^{(0,0)}\,,\\
%%%%%%%%%%%%%%%%%%%%%%%%
\bd{B}_3^{(0,1)}\ssb\ =&\ \bd{B}_3^{(0,1)}\mid^{(4,2)} + s \bd{B}_3^{(0,1)}\mid^{(2,2)}
+ \bar{s} \bd{B}_3^{(0,1)}\mid^{(4,0)} 
\nonumber\\
&\
+ s^2 \bd{B}_3^{(0,1)}\mid^{(0,2)}
+ s\bar{s} \bd{B}_3^{(0,1)}\mid^{(2,0)}
 + s^2\bar{s} \bd{B}_3^{(0,1)}\mid^{(0,0)}\,,\\
%%%%%%%%%%%%%%%%%%%%%%%%
\bd{B}_3^{(2,0)}\ssb\ =&\ \bd{B}_3^{(2,0)}\mid^{(0,4)} + \bar{s} \bd{B}_3^{(2,0)}\mid^{(0,2)}
+ \bar{s}^2\bd{B}_3^{(2,0)}\mid^{(0,0)}\,,\\
%%%%%%%%%%%%%%%%%%%%%%%%
\bd{B}_3^{(1,1)}\ssb\ =&\ \bd{B}_3^{(1,1)}\mid^{(2,2)} + s \bd{B}_3^{(1,1)}\mid^{(0,2)}
+ \bar{s} \bd{B}_3^{(1,1)}\mid^{(2,0)} + s\bar{s} \bd{B}_3^{(1,1)}\mid^{(0,0)}\,,\\
%%%%%%%%%%%%%%%%%%%%%%
\bd{B}_3^{(0,2)}\ssb\ =&\ \bd{B}_3^{(0,2)}\mid^{(4,0)} + s \bd{B}_3^{(0,2)}\mid^{(2,0)}
+ s^2\bd{B}_3^{(0,2)}\mid^{(0,0)}\,,\\
%%%%%%%%%%%%%%%%%%%%%%%%
\bd{B}_3^{(2,1)}\ssb\ =&\ \bd{B}_3^{(2,1)}\mid^{(0,2)} + \bar{s} \bd{B}_3^{(2,1)}\mid^{(0,0)}\,,\\
%%%%%%%%%%%%%%%%%%%%%%%% 
\bd{B}_3^{(1,2)}\ssb\ =&\ \bd{B}_3^{(1,2)}\mid^{(2,0)} + s \bd{B}_3 ^{(1,2)}\mid^{(0,0)}\,,\\
%%%%%%%%%%%%%%%%%%%%%%%%
\bd{B}_3^{(2,2)}\ssb\ =&\ \bd{B}_3^{(2,2)}\mid^{(0,0)}\,,
%%%%%%%%%%%%%%%%%%%%%%%%%%%%%%%%%%%%%%%%%%%%%%%
\end{align}
and
\begin{align}
%%%%%%%%%%%%%%%%%%%%%%%%
\bd{\lambda}_3^{(1,0)}\ssb\ =&\ \bd{\lambda}_3^{(1,0)}\mid^{(2,4)} + s \bd{\lambda}_3^{(1,0)}\mid^{(0,4)}
+ \bar{s} \bd{\lambda}_3^{(1,0)}\mid^{(2,2)} 
\nonumber\\
&\
+ s\bar{s} \bd{\lambda}_3^{(1,0)}\mid^{(0,2)}
+ \bar{s}^2 \bd{\lambda}_3^{(1,0)}\mid^{(2,0)} + s\bar{s}^2 \bd{\lambda}_3^{(1,0)}\mid^{(0,0)}\,,\\
%%%%%%%%%%%%%%%%%%%%%%
\bd{\lambda}^{(2,0)}\ssb\ =&\ \bd{\lambda}_3^{(2,0)}\mid^{(0,4)} + \bar{s} \bd{\lambda}_3^{(2,0)}\mid^{(0,2)}
+ \bar{s}^2 \bd{\lambda}_3^{(2,0)}\mid^{(0,0)}\,,\\
%%%%%%%%%%%%%%%%%%%%%%
\bd{\lambda}_3^{(1,1)}\ssb\ =&\ \bd{\lambda}_3^{(1,1)}\mid^{(2,2)} + s \bd{\lambda}_3^{(1,1)}\mid^{(0,2)}
+ \bar{s} \bd{\lambda}_3^{(1,1)}\mid^{(2,0)} + s\bar{s} \bd{\lambda}_3^{(1,1)}\mid^{(0,0)}\,,\\
%%%%%%%%%%%%%%%%%%%%%%
\bd{\lambda}_3^{(2,1)}\ssb\ =&\ \bd{\lambda}_3^{(2,1)}\mid^{(0,2)} + \bar{s} \bd{\lambda}_3^{(2,1)}\mid^{(0,0)}\,,\\
%%%%%%%%%%%%%%%%%%%%%%
\bd{\lambda}_3^{(1,2)}\ssb\ =&\ \bd{\lambda}_3^{(1,2)}\mid^{(2,0)} + s \bd{\lambda}_3^{(1,2)}\mid^{(0,0)}\,,\\
%%%%%%%%%%%%%%%%%%%%%%
\bd{\lambda}_3^{(2,2)}\ssb\ =&\ \bd{\lambda}_3^{(2,2)}\mid^{(0,0)}\,,\\
%%%%%%%%%%%%%%%%%%%%%%%%%%%%%%%%%%%%%%%%%%%%%%%
\bar{\bd{\lambda}}_3^{(0,1)}\ssb\ =&\ \bar{\bd{\lambda}}_3^{(0,1)}\mid^{(4,2)} 
+ s \bar{\bd{\lambda}}_3^{(0,1)}\mid^{(2,2)}
+ \bar{s} \bar{\bd{\lambda}}_3^{(0,1)}\mid^{(4,0)} 
\nonumber\\
&\
+ s^2 \bar{\bd{\lambda}}_3^{(0,1)}\mid^{(0,2)} + s\bar{s} \bar{\bd{\lambda}}_3^{(0,1)}\mid^{(2,0)}
+ s\bar{s}^2 \bar{\bd{\lambda}}_3^{(0,1)}\mid^{(0,0)}\,,\\
%%%%%%%%%%%%%%%%%%%%%%
\bar{\bd{\lambda}}_3^{(1,1)}\ssb\ =&\ \bar{\bd{\lambda}}_3^{(1,1)}\mid^{(2,2)} 
+ s \bar{\bd{\lambda}}_3^{(1,1)}\mid^{(0,2)}
+ \bar{s} \bar{\bd{\lambda}}_3^{(1,1)}\mid^{(2,0)} + s\bar{s} \bar{\bd{\lambda}}_3^{(1,1)}\mid^{(0,0)}\,,\\
%%%%%%%%%%%%%%%%%%%%%%
\bar{\bd{\lambda}}^{(0,2)}\ssb\ =&\ \bar{\bd{\lambda}}_3^{(0,2)}\mid^{(4,0)} 
+ s \bar{\bd{\lambda}}_3^{(0,2)}\mid^{(2,0)}
+ s^2 \bar{\bd{\lambda}}_3^{(0,2)}\mid^{(0,0)}\,,\\
%%%%%%%%%%%%%%%%%%%%%%
\bar{\bd{\lambda}}_3^{(2,1)}\ssb\ =&\ \bar{\bd{\lambda}}_3^{(2,1)}\mid^{(0,2)} 
+ \bar{s} \bar{\bd{\lambda}}_3^{(2,1)}\mid^{(0,0)}\,,\\
%%%%%%%%%%%%%%%%%%%%%%
\bar{\bd{\lambda}}_3^{(1,2)}\ssb\ =&\ \bar{\bd{\lambda}}_3^{(1,2)}\mid^{(2,0)} 
+ s \bar{\bd{\lambda}}_3^{(1,2)}\mid^{(0,0)}\,,\\
%%%%%%%%%%%%%%%%%%%%%%
\bar{\bd{\lambda}}_3^{(2,2)}\ssb\ =&\ \bar{\bd{\lambda}}_3^{(2,2)}\mid^{(0,0)}\,.
\end{align}
The flow of each product is shown in Figs.~\ref{3string comp1} and \ref{3string comp2}.
The 3-products at the end of flows are used to make the $L_\infty$ structure for 
writing down the action, which are those with no picture number deficit.

\end{document}